\documentclass[12pt,preprint]{emulateapj}

\usepackage{appendix}
\usepackage{graphicx}
\usepackage{amssymb}
\usepackage{epstopdf}
\usepackage{subfig}

\newcommand{\ba}{{\mbox{\boldmath$a$}}}
\newcommand{\bb}{{\mbox{\boldmath$b$}}}

\newcommand{\bp}{{\mbox{\boldmath$p$}}}

\newcommand{\br}{{\mbox{\boldmath$r$}}}

\newcommand{\bu}{{\mbox{\boldmath$u$}}}
\newcommand{\bvv}{{\mbox{\boldmath$v$}}}

\newcommand{\bx}{{\mbox{\boldmath$x$}}}
\newcommand{\by}{{\mbox{\boldmath$y$}}}

\newcommand{\bz}{{\mbox{\boldmath$z$}}}

\newcommand{\bB}{{\mbox{\boldmath$B$}}}

\newcommand{\bE}{{\mbox{\boldmath$E$}}}

\newcommand{\bI}{{\mbox{\boldmath$I$}}}

\newcommand{\bV}{{\mbox{\boldmath$V$}}}

\newcommand{\om}{{\mbox{$\omega$}}}

\newcommand{\bnabla}{\mbox{\boldmath$\nabla$}}
\newcommand{\bkappa}{\mbox{\boldmath$\kappa$}}

\newcommand{\Dpar}{{\mbox{$D_{\parallel}$}}}
\newcommand{\Dperp}{{\mbox{$D_{\perp}$}}}
\newcommand{\kpar}{{\mbox{$\kappa_{\parallel}$}}}
\newcommand{\kperp}{{\mbox{$\kappa_{\perp}$}}}
\def  \be   {\begin{equation}}
\def  \ee   {\end{equation}}
\def  \beq  {\begin{eqnarray}}
\def  \eeq  {\end{eqnarray}}

\begin{document}

\shorttitle{Cross-field Cosmic Ray Transport}
\shortauthors{Desiati \& Zweibel}

\title{The Transport of Cosmic Rays Across Magnetic Fieldlines}

\author{Paolo Desiati}
\affil{Wisconsin IceCube Particle Astrophysics Center (WIPAC) and Department of Astronomy, University of Wisconsin-Madison,
222 W. Washington Ave, Madison, WI 53703; {\sf desiati@wipac.wisc.edu}}
\author{Ellen G. Zweibel}
\affil{Departments of Astronomy and Physics, University of Wisconsin-Madison, and Center for Magnetic Self-Organization,  
475 N. Charter St., Madison, WI 53706; {\sf zweibel@astro.wisc.edu}}


\begin{abstract}
The long residence times and small anisotropies of cosmic rays suggest that they are well confined and well scattered by the Galactic magnetic field. Due to the disklike shape of the confinement volume, transport in the vertical direction, perpendicular to the mean Galactic magnetic field, is key to cosmic ray escape. It has long been recognized that this vertical transport depends both on the vertical component of the fieldlines themselves and on the extent to which the cosmic rays are tied to the fieldlines. In this paper we use magnetic fields with
very simple spatial and temporal structure to isolate some important features of cross fieldline transport. We show that even simple magnetic nonuniformities combined with pitch angle scattering can enhance cross fieldline transport by several orders of magnitude, while pitch angle scattering is unnecessary for enhanced transport if the field is chaotic. Nevertheless, perpendicular transport is much less than
parallel transport in all the cases we study. We apply the results to confinement of cosmic rays in the Fermi Bubbles.
\end{abstract}


\section{Introduction}\label{sec:introduction}

It is well accepted that cosmic rays below $\sim 10^{18}$ eV are confined to the Milky Way by the Galactic magnetic field. The
confinement volume is a thick disk, suggesting that the primary direction of escape is perpendicular to the Galactic plane. Furthermore, the distribution of cosmic ray arrival directions at the Earth is
nearly isotropic.
These features, together with observations of the Galactic magnetic field which show it to be oriented preferentially parallel to the Galactic plane but with a substantial randomly oriented component,
 have led to a picture in which cosmic rays are isotropized by scattering from small amplitude, short wavelength magnetic fluctuations and achieve their vertical displacement both by following
the magnetic fieldlines as they snake their way out of the Galactic plane and by transport across the field itself (see \cite{hillas_2006, aharonian_2012, zweibel_2013} for some recent reviews). 

This spatial transport with respect to the background medium (as opposed to, e.g., advective transport by a Galactic wind) is frequently described by a diffusion tensor $\bkappa\equiv\kappa_{\perp}\bI - (\kappa_{\perp}-\kappa_{\parallel})\bB\bB/B^2$, where $\bI$ is the unit tensor and
$\bB$ is the {\textit{mean}} magnetic field. 
Simple parameterizations of the diffusion tensor have led to
many successful descriptions of cosmic ray properties (e.g. \cite{strong_1998, strong_2010, vladimirov_2011}).

Estimating $\kappa_{\perp}$ and $\kappa_{\parallel}$, 
however, and even establishing the circumstances under which transport is diffusive ($\langle\Delta x^2\rangle\propto\Delta t$),
as opposed to
subdiffusive ($\langle\Delta x^2\rangle\propto\Delta t^{\alpha}$; $\alpha < 1$), or superdiffusive ($\langle\Delta x^2\rangle\propto\Delta t^{\alpha}$; $\alpha > 1$), has proven to be difficult. 
 Transport results from the interplay of many effects: the large scale magnetic field geometry, the properties of the turbulence, the rate of scattering by small scale fluctuations, and
the extent to which magnetic fieldlines diverge on scales less than the particle gyroradius. It is generally agreed that if the particles are tied to the fieldlines, the maximum perpendicular diffusion rate is set by the rate of perpendicular fieldline wandering, scaled by particle velocity (Fieldline Random Walk, or FLRW), and that parallel scattering reduces $\kappa_{\perp}$ below its FLRW value \citep{jokipii_1966, jokipii_1969, rechester_1978, giacalone_1999, minnie_2009}.
For a recent review of the subject, see \cite{shalchi_2009}.

In this paper, we address a different problem: the rate at which cosmic rays cross the {\textit{exact}} field, rather than the {\textit{average}} field. This
problem has received relatively less attention (see \cite{barge_1984} however). While for cosmic ray confinement in the Galaxy it is primarily vertical transport (i.e. perpendicular to the mean magnetic field) which is important,  transport across the exact field matters for other problems. Confinement of cosmic rays to
bubbles blown by AGN, which presumably are bounded by magnetic surfaces is one such problem. The Fermi Bubbles \citep{finkbeiner_2004, dobler_2010} - which may have been created by sporadic activity in the Galactic Center - are a local example. Numerical simulations of bubble
evolution with perpendicular cosmic ray diffusion suppressed yield surface
brightness profiles which agree quite well with observations, while surface brightness profiles based on simulations with $\kappa_{\perp}=\kappa_{\parallel}$ are more diffuse than observed \citep{yang_2012}.

Cross fieldline transport is easily estimated in two extreme cases: drifting due to large scale structure, and scattering by small scale structure. Drifts occur either when the direction or strength of the magnetic field varies in space or when electrical or gravitational acceleration $\ba$ acts on the particles. The drift velocity $v_D$ produced by magnetic field inhomogeneity on scale $L$ can be estimated as $v_D/v \sim r_g/L$, where $v$ is the particle speed and $r_g$ is the particle Larmor radius. The drifts induced by nonmagnetic forces are of order $a_{\perp}/\om_{g}$,
where $a_{\perp}$ is the component of nonmagnetic acceleration normal to $\bB$ and 
$\om_{g}$ 
is the relativistic gyrofrequency. If one adopts values of $L$ and $a_{\perp}$ characteristic of global properties of the
Galaxy, then the drifts are too small to have anything to do with cosmic ray escape, except for very high energy cosmic rays. However, there is a certain amount of random drifting due to moderate scale
turbulence.

Scattering by small ($r_g$ scale) fluctuations is generally thought to be the mechanism responsible for parallel diffusion, but also produces perpendicular diffusion.
Every pitch angle scattering event $\delta\theta\sim \delta B/B$, where $\delta B/B$ is the fluctuation relative amplitude, is accompanied by a cross fieldline displacement of order $\delta r\sim r_g\delta\theta$. The corresponding perpendicular diffusion coefficient $\kappa_{\perp}\sim v r_g$, resulting in a vertical escape time of order $L^2/(v r_g)$, is similar to what is estimated from drifts and, again, too small to have anything to do with the escape of GeV cosmic rays, where the bulk of the cosmic ray distribution lies.

If magnetic fieldlines separated by less than a fiducial gyroradius wander away from each other, particles do not follow fieldlines. This case is difficult to treat analytically, but has been studied numerically \citep{jokipii_1973, qin_2002, hauff_2010, lazarianyan_2013}, together with its effects on diffusion across the mean field.

A realistic numerical study of cosmic ray propagation based on integration of particle orbits would include both the large scale geometry of the Galactic magnetic field and the small scale fluctuations which provide pitch angle scattering and possibly second order Fermi acceleration. Such a magnetic field model would have to span at least 10 orders of magnitude in lengthscale, while current numerical
simulations of interstellar magnetized turbulence rarely capture more than two orders of magnitude in scale. And even if one prescribed a magnetic field model with a realistic range of scales, accurately integrating large enough numbers of particle orbits over long times and distances required for a statistically representative sample would be prohibitively expensive.

In this study of particle transport, we address the range of scales problem by integrating particle orbits in magnetic field structures much larger than a larmor radius with pitch angle scattering by gyroradius scale fluctuations modeled as a random process. A similar approach was followed in \cite{neuer_2006a, neuer_2006b}, but their random process, which is intended to model collisions, does not conserve energy, while ours, which is intended to model pitch angle scattering by nearly static fluctuations, does.
This dual approach allows us to probe the effects of fieldline geometry and variation of the field on large scales while retaining pitch angle diffusion, and at the same time avoiding the need to integrate over a large dynamic range in scales of magnetic structure. We find that in many cases diffusive behavior is recovered, but that the diffusivities do not depend on fluctuation amplitudes according to the prescription derived from quasilinear theory. 

In \S\ref{sec:setup} we introduce the basic magnetic field models we use, discuss the implementation of pitch angle scattering, show how we measure diffusion, and establish upper and lower bounds on the cross
field diffusivity. In \S\ref{sec:results} we present the results for the different field models and discuss their physical basis. Section \ref{sec:discussion} is a discussion and comparison with other results and \S\ref{sec:conclusions} is a summary.

\section{Setup of the Problem}\label{sec:setup}

The particle trajectories in a generic electromagnetic field are computed by numerically integrating the equation of motion
\begin{equation}\label{momentum}
\frac{d\bp}{dt}=q\left(\bE+\frac{\bvv\times\bB}{c}\right)
\end{equation}
for prescribed electric and magnetic fields $\bE(\bx, t)$, $\bB(\bx, t)$, where $\bp = m\gamma \bvv$ is the relativistic momentum of a particle with rest mass $m$. In the remainder of the paper, we work with a dimensionless version of eqn. (\ref{momentum}). We introduce a fiducial magnetic field strength $B_0$, which in practice will be the magnitude of the mean, or guide field. We then adopt dimensionless magnetic and electric fields $\hat\bB\equiv\bB/B_0$, $\hat\bE\equiv\bE/B_0$. We introduce the fiducial proton gyrofrequency $\om_0\equiv eB_0/m_pc$ and replace time $t$ by $s\equiv\om_0t$. We introduce a fiducial gyroradius $r_0\equiv c/\om_0$ and express lengths in units of $r_0$; $\hat\bx\equiv\bx/r_0$. We express momentum in units of $mc$; $\hat\bp\equiv\bp/mc$. The particle velocity $\bvv$ is related to $\hat\bp$ by $\bvv = \hat\bp/\sqrt{1+\hat p^2}$ and the particle Lorentz factor $\gamma = \sqrt{1+\hat p^2}$. In these units the dimensionless particle gyroradius is $\hat r_g = \hat p_{\perp}$, and the dimensionless gyrofrequency is $\hat \omega_g = 1/\gamma$. We denote normalized variables with hats. 
With these definitions and substitutions, the equation of motion can be re-written as
\begin{equation}\label{dimlessmomentum}
\left\{
\begin{array}{lcl}
\frac{d\hat\bp}{ds} & = & \hat\bE+\frac{\hat\bp}{\sqrt{1+\hat p^2}}\times\hat\bB \\
\frac{d\hat \bx}{ds}       & = & \frac{\hat\bp}{\sqrt{1+\hat p^2}}.
\end{array}
\right.
\end{equation}

\subsection{Magnetic Field Configurations}
\label{ssec:bfields}

 We have integrated test particle orbits for three simple magnetic field configurations: a uniform magnetic field, and two types of simple cellular magnetic field structures, one of which is time dependent.
 These magnetic field configurations are used to study 
cross fieldline propagation from the combined effect of pitch angle scattering from gyroscale magnetic perturbations and nontrivial large scale magnetic geometry. These fields are 2.5D, with three spatial components and an ignorable coordinate, $z$.

The uniform magnetic field is simply $\hat \bB = \hat \bz$, i.e. with no explicit perpendicular perturbation (apart from the gyroscale perturbations effectively accounted for by pitch angle scattering and described in \S\ref{ssec:sca}). We study this case to test our code against well established predictions from quasilinear theory and to determine baseline diffusion rates. This is intended as a benchmark to compare with cases where the effects of a large scale magnetic field structure are considered. The other two models can be represented as a superposition of the guide field $\hat \bz$ and a perpendicular component
\begin{equation}\label{cellular}
\hat\bB=\hat \bz +\bnabla\hat A(x,y)\times\hat \bz.
\end{equation}
The so-called cellular magnetic field is represented by the magnetic potential
\begin{equation}\label{cellularA}
\hat A_C = -\frac{\epsilon}{\hat k}\,\sin{\hat k} {\hat x}\, \sin{\hat k} {\hat y},
\end{equation}
with $\epsilon$ the amplitude (assumed to be small) of the perturbation relative to the guide field strength and $\hat k=\frac{2\pi}{\hat L}$, where $\hat L$ sets the geometrical scale of the perturbation. Eqns. (\ref{cellular}) and (\ref{cellularA}) describe a periodic array of cells with helical magnetic fieldlines spiraling around the points $(x,y) = (\frac{(2n+1)\pi}{2},\frac{(2m+1)\pi}{2})$ and cell boundaries, or separatrix lines, at $x=n\pi$, $y=m\pi$ (for integer $n$ and $m$).
%
%

The third magnetic field model is a time dependent cellular field of the form (\ref{cellular}) but with vector potential
\begin{equation}\label{AGP}
\hat A_{GP}=\frac{\epsilon}{\hat k}\left[\cos{(\hat k\hat x+\hat \lambda\cos{\hat \om s})}+\sin{(\hat k\hat y+\hat \lambda\sin{\hat \om s})}\right],
\end{equation}
%
%
Eqn. (\ref{AGP}) describes a periodic array of cells (rotated by $\pi/4$ with respect to the cells described by $\hat A_C$)
which translates back and forth by a scaled distance $\pm\hat \lambda/\hat k$ with period $\hat \tau = 2\pi/\hat \om$. We call this the Galloway-Proctor (GP) field, because the magnetic fieldlines corresponding to $\hat A_{GP}$ have the same structure as the flow 
shown by \cite{galloway_1992} to amplify magnetic fields exponentially fast in the 
limit of large magnetic Reynolds number (i.e. it is a fast dynamo). 

The GP {\it flow} is chaotic in the sense that neighboring streamlines separate exponentially. The rates of separation, or Lyapunov exponents, have been computed by \cite{brummell_2001, heitsch_2004}, and are found to
peak at $\hat\lambda\sim 0.25$; we therefore refer to $\hat\lambda$ as the complexity parameter. The GP {\it flow} is known to rapidly mix  passive scalar quantities, i.e properties or particles which are assumed not to affect the underlying flow
 \citep{brummell_2001, heitsch_2004}. 

The efficient mixing properties of the GP {\it flow} suggest that particle transport in magnetic fields with the same structure  would likewise be efficient. This is
because a particle moving exactly along the field with speed $u$ and velocity $\bu=u\bB/B$ would simply follow the GP {\it flow}. We will see in \S 3.3 that 
transport by the GP magnetic field is indeed faster than transport in the cellular field.


Because $\hat A_{GP}$ is time dependent, there is an electric field in this model
\begin{equation}\label{EGP}
\hat \bE_{GP}=-\frac{\hat \bz}{c}\frac{\partial \hat A_{GP}}{\partial t}.
\end{equation}
We can think of $\bE_{GP}$ as the inductive electric field produced by a flow $\bV$, where
\begin{equation}\label{ExB}
\bV=-\frac{\partial A}{\partial t}\frac{\bnabla A}{\vert\nabla A\vert^2}.
\end{equation}
The amplitude of $V$ is of order $c\hat\lambda\hat\om/\hat k$ and will turn out to range from a few km/s to a few thousand km/s for the parameters used in this study.

For both the cellular and the GP {\it flow} magnetic field configurations a spatial scale of $\hat L=200$ was used in most cases, meaning that $\hat k = \pi/100$ and each cell has a side of length $100$ (we also did a few cases with larger $\hat L$ to test the $\hat L$ dependence of the results). 
The amplitude parameter $\epsilon$ was chosen to be $\epsilon = 10^{-3}, 10^{-2}, 10^{-1}$.



\subsection{Numerical Approach}
Integration of particle trajectories in magnetic fields is very sensitive to numerical accuracy. Although most of the numerical results reported here involve random pitch angle scattering, which ``resets" properties of the orbits so that numerical errors do not accumulate, we sought robust integration methods and tested them extensively. In the Appendix a description of the integration methods used in this work and of their accuracy level is discussed in detail. We stress that the accuracy of numerical solutions of particle trajectory integration for a given magnetic system and particle energy is an essential component of such studies. 



%
%
\subsection{Scattering}
\label{ssec:sca}

Cosmic rays are
pitch angle scattered by gyroscale magnetic field fluctuations, which may be hydromagnetic waves. The scattering frequency $\nu$ can be expressed in terms of the relativistic gyrofrequency $\om_g$ and relative magnetic perturbation amplitude $\delta\hat B\equiv \frac{\delta B}{B_0}$ as \citep{kulsrud_2005}
\begin{equation}\label{nu}
\nu = \frac{\pi}{2}\om_g\delta\hat B^2.
\end{equation}
Scattering produces diffusion in pitch angle $\mu\equiv\bp\cdot\bB/pB$ with diffusion coefficient
\begin{equation}\label{dmm}
D_{\mu\mu}\equiv\frac{\langle(\Delta\mu)^2\rangle}{2\Delta t}=\frac{\nu (1-\mu^2)}{2}.
\end{equation}

 We simulate pitch angle scattering by a random process. At time intervals
$\Delta t = \frac{2\pi}{\om_g} = \frac{2\pi}{\om_0} \sqrt{1+\hat p^2}$, corresponding to one relativistic gyroperiod at field strength $B_0$, we choose a random angle $\psi$ and increment $\mu$ by an amount $\Delta\mu=\sqrt{1-\mu^2}\epsilon_f\sin{\psi}$, where $\epsilon_f$ is a small constant number. The corresponding Fokker Planck diffusion coefficient is
\begin{equation}\label{dmu2dt}
\frac{\langle (\Delta\mu)^2\rangle}{2\Delta t}=\frac{\om_g}{8\pi} (1-\mu^2)\epsilon_f^2,
\end{equation}
where we used the relationship $\langle sin^2 \psi \rangle = \frac{1}{2}$. So our scattering frequency agrees with eqns. (\ref{nu}) and (\ref{dmm}) if we set $\epsilon_f^2={2\pi^2}\delta\hat B^2$. Eqn. (\ref{dmu2dt}) is valid as long as $\delta\hat B \ll 1$.

We assume scattering does not change particle energy. This is a good approximation for scattering by hydromagnetic waves with Alfv\'en velocity $v_A$, for which the momentum diffusion coefficient $D_{pp}\sim (v_A/c)^2 D_{\mu\mu}\ll D_{\mu\mu}$ under typical interstellar conditions. The change in vector momentum corresponding to each scattering is then
$\Delta\hat p_{\parallel} = \hat p \Delta\mu$,
$\Delta\hat p_{\perp} = -\hat p\mu \Delta\mu/ \sqrt{1-\mu^2}$
to lowest order in $\epsilon_f$. These expressions are well behaved as $\mu\rightarrow\pm 1$ and do not lead to singular behavior. 

We assume the instantaneous particle positions are unchanged by the scattering. The position of the guide center $r_{gc}$, however, is changed by an amount of order $r_g \delta\theta$, which can be found by integrating the equations of motion
\begin{equation}\label{deltargc}
\Delta\br_{gc}=\frac{\Delta\bvv\times\bB}{\om_g B},
\end{equation}
with $\bvv$ the particle velocity. This leads to cross fieldline diffusion.

\subsection{Spatial diffusion}\label{subsec:spatialdiffusion}

The running spatial diffusion tensor $D_{ij}$ for an ensemble of $N$ particles is defined as
\begin{equation}\label{runningD}
D_{ij}(t)\equiv \frac{1}{2N} \sum_{n=1}^{N} \frac{[x_{i,n}(t)-x_{i,n}(0)][x_{j,n}(t)-x_{j,n}(0)]}{t},
\end{equation}
where $x_{i,n}(0)$ denotes the $i^{th}$ component of initial position of the $n^{th}$ particle \citep{fraschetti_2012}.
In all the cases we examined, the particle distribution is gyrotropic and parallel and perpendicular motions are decoupled. This implies that $D_{ij}$ is diagonal, with two identical perpendicular components $D_{xx}=D_{yy}=\Dperp$ and a parallel component $\Dpar$.
In the absence of scattering the perpendicular displacements of magnetized particles are bounded, so after one gyroperiod $\Dperp$ decays as $t^{-1}$. For the magnetic field models studied here, there is a dominant unidirectional guide field, so in the absence of scattering $\Dpar$ increases as $t$. If the particle motion is a random walk, with the rms displacement increasing as $t^{1/2}$, $D_{ij}$ becomes independent of time, which corresponds to the diffusion regime for the ensemble of particles. Eqn. (\ref{runningD}) is readily computed for large ensembles of particles, and all measured diffusion coefficients reported here are calculated from this formula. 
When the running diffusion coefficient levels off at a stable value, we denote that value by $\kappa$.

The quantities defined in eqn. (\ref{runningD}) measure displacement in space, not displacement with respect to the magnetic field. In order to describe the true cross fieldline diffusivity, we assign a guiding
center $\bx_{gc}$ to a particle at position $\bx$ by solving an analog of eqn. (\ref{deltargc})
\begin{equation}\label{xgc}
\bx_{gc}=\bx+\frac{\bvv\times\hat\bb}{\om_g}.
\end{equation}
We integrate the magnetic fieldline that passes through the initial position $\bx_{gc}(0)$ up to the value of $z$ where the particle is, and denote its position in the $(x,y)$ plane by $\bx_f$. The displacement
of a particle guiding center relative to its original fieldline is then 
\begin{equation}\label{deltax}
\Delta\bx\equiv\bx_{gc}-\bx_f.
\end{equation}
We then define
the running cross-field diffusion coefficient $D_{ij}^c$ by
\begin{equation}\label{runningDcross}
D_{ij}^c\equiv\frac{1}{2N}\sum_{n=1}^N\frac{\Delta x_i(t)\Delta x_j(t)}{t}
\end{equation}
(note that $\Delta\bx(0)\equiv 0$).

This procedure is time consuming to apply. A simpler, but less rigorous procedure, is to compare two runs, one with scattering and one without. Denoting the positions of a particle with and without scattering by $\bx_s$, $\bx_{ns}$, we then form the quantity $\tilde\Delta\bx\equiv\bx_s-\bx_{ns}$ and use the $\tilde\Delta\bx$ in eqn. (\ref{runningDcross}). Both these methods assume implicitly that in the absence of scattering,
the particles follow well defined gyro-orbits and are well tied to fieldlines. These assumptions break down in the absence of a strong guide field that is nearly uniform over the particle orbits.

We have found that although the $D_{\perp}$ and $D_{\perp}^c$ differ at early times, when particles undergo perpendicular displacement simply by following the fieldlines as they wind around the cell axis,
they converge to the same value at late times, when particles have migrated out of their original cells. In fact, we can say with certainty, that all horizontal displacements greater than $L/\sqrt{2}$ are due
primarily to cross-field transport. Therefore, we present results for $\kappa$ computed using $D$ instead of $D^c$, due to the ease of doing so. 

Pitch angle scattering induces a decorrelation of particle velocities with respect to the unperturbed trajectories.
This can be measured by computing the pitch angle correlation function
\begin{equation}
\langle\mu(t) \mu(0)\rangle = \frac{\sum_{i=1}^{N}\mu_i(t)\mu_i(0)}{\sum_{i=1}^{N}\mu_i(0)^2}.
\end{equation}
%
 By definition, the pitch angle correlation function equals unity initially, and will be seen to drop nearly to zero on a timescale of order the scattering time $\tau_{sca}\equiv\nu^{-1}$ as the particles decorrelate. 

It was shown by \cite{kulsrud_1969} that pitch angle scattering leads to spatial diffusion as long as it is frequent compared to global dynamical times, or when the mean free path $\lambda_{\parallel}\equiv v/\nu$ is short compared with global length scales. The parallel diffusion coefficient $\kpar$ is defined to be
\begin{equation}\label{kpar}
\kpar = \langle\frac{ v^2_{\parallel}}{\nu}\rangle = v^2 \frac{\int_{-1}^{1}f(\mu)\frac{ \mu^2}{\nu} d\mu}{\int_{-1}^{1} d\mu} = \frac{v^2}{3\nu},
\end{equation}
where $f(\mu)$ is the pitch angle distribution function at fixed momentum and the last equality holds for an isotropic distribution function in pitch angle and a scattering frequency independent of $\mu$.

The perpendicular spatial diffusion coefficient follows from eqn. (\ref{deltargc}) and is given by \cite{parker_1965} and \cite{forman_1975}. When $\lambda_{\parallel}/r_g\gg 1$ (which corresponds to $\delta\hat B\ll 1$), we have
\begin{equation}\label{kperp}
\kperp=\kappa_{\parallel}\frac{r_g^2}{\lambda_{\parallel}^2}=\frac{r_g^2\nu}{3},
\end{equation}
where as in eqn. (\ref{kpar}) the last equality holds for an isotropic distribution and a $\mu$-independent scattering coefficient. Both coefficients are proportional to $v\hat p$, the different scale being determined by $\delta\hat B$. Comparing eqns. (\ref{kpar}) and (\ref{kperp}) and using eqn. (\ref{nu}) we see that
$\kperp/\kpar = (\frac{\pi}{2})^2 (\delta\hat B)^4 \ll 1$ for small angle scattering.

In the remainder of the paper we will use $\Dpar$ and $\Dperp$ for the running diffusion coefficients calculated from our simulations, and use $\kpar$ and $\kperp$ for the converged expressions. We non-dimensionalize all diffusion coefficients by expressing them in units of $r_0^2\,\om_0$.

\section{Results}\label{sec:results}
Salient numerical data from some of the runs discussed in this paper are given in Tables \ref{tab:uniruns}, \ref{tab:cellruns} and \ref{tab:gpruns}.
\subsection{The Uniform Field}\label{ssec:uniform}
Cross-field transport by small amplitude, random pitch angle scattering in a uniform magnetic field is well understood. We include this case to demonstrate that our numerical method successfully reproduces  results obtained previously from quasilinear theory and  to establish a baseline level of cross fieldline diffusion which can be compared to cross fieldline transport in less trivial geometries.

\begin{figure*}[!ht]
\begin{center}
\subfloat[\label{fig:rundiffuniforma}]{%
  \includegraphics[width=\columnwidth]{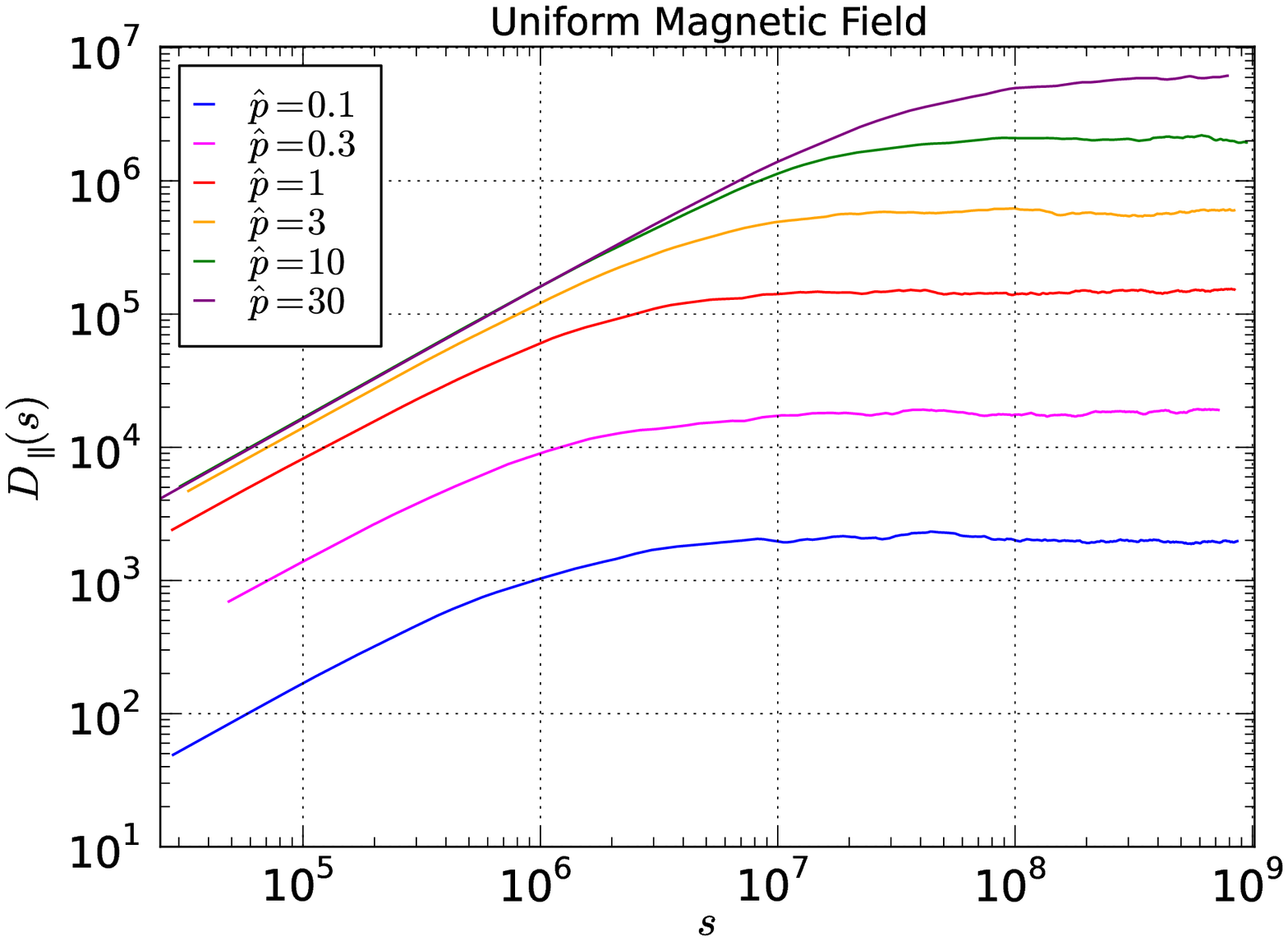}
  }
\subfloat[\label{fig:rundiffuniformb}]{%
  \includegraphics[width=\columnwidth]{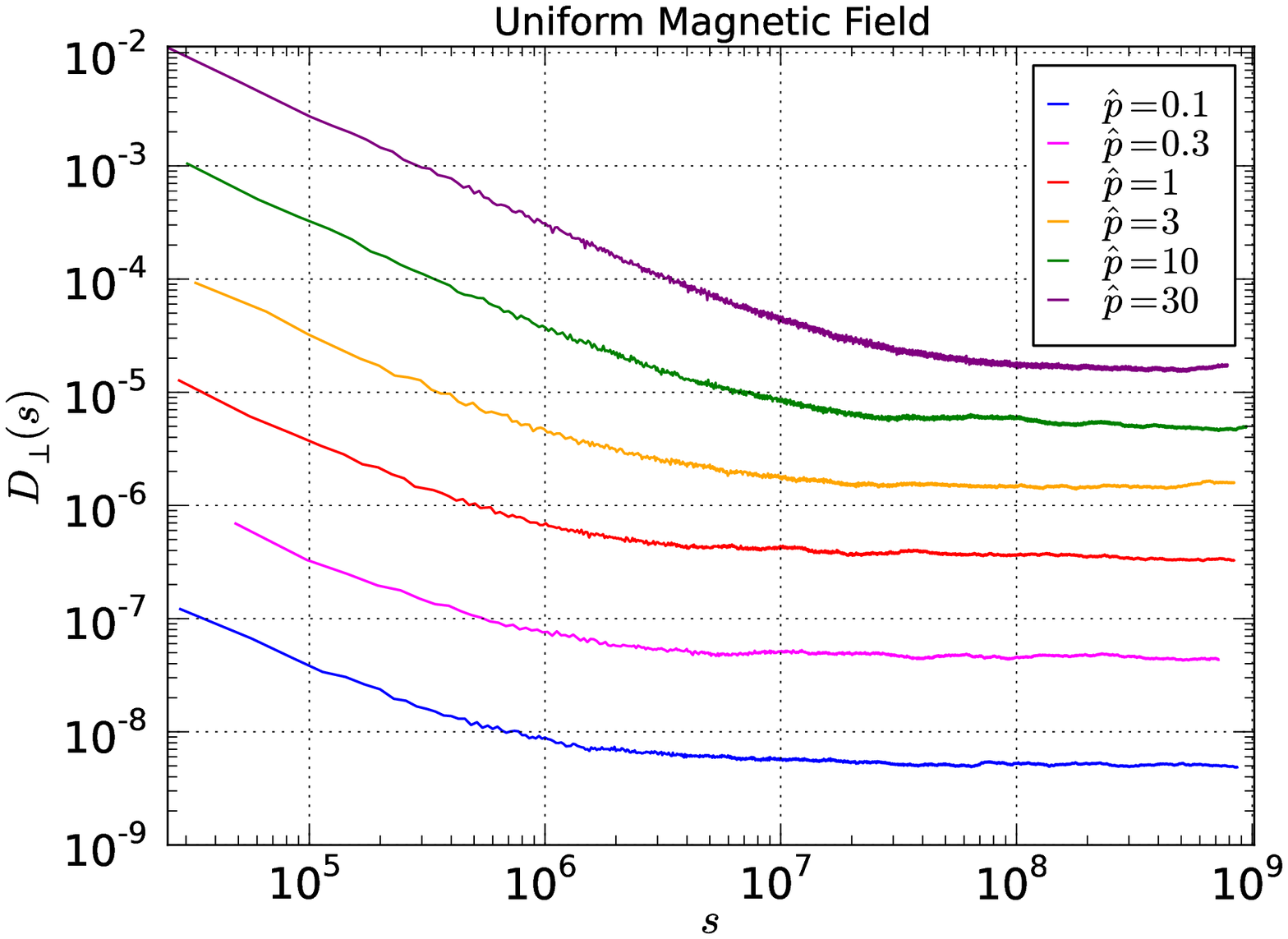}
  }

\subfloat[\label{fig:rundiffuniformc}]{%
  \includegraphics[width=\columnwidth]{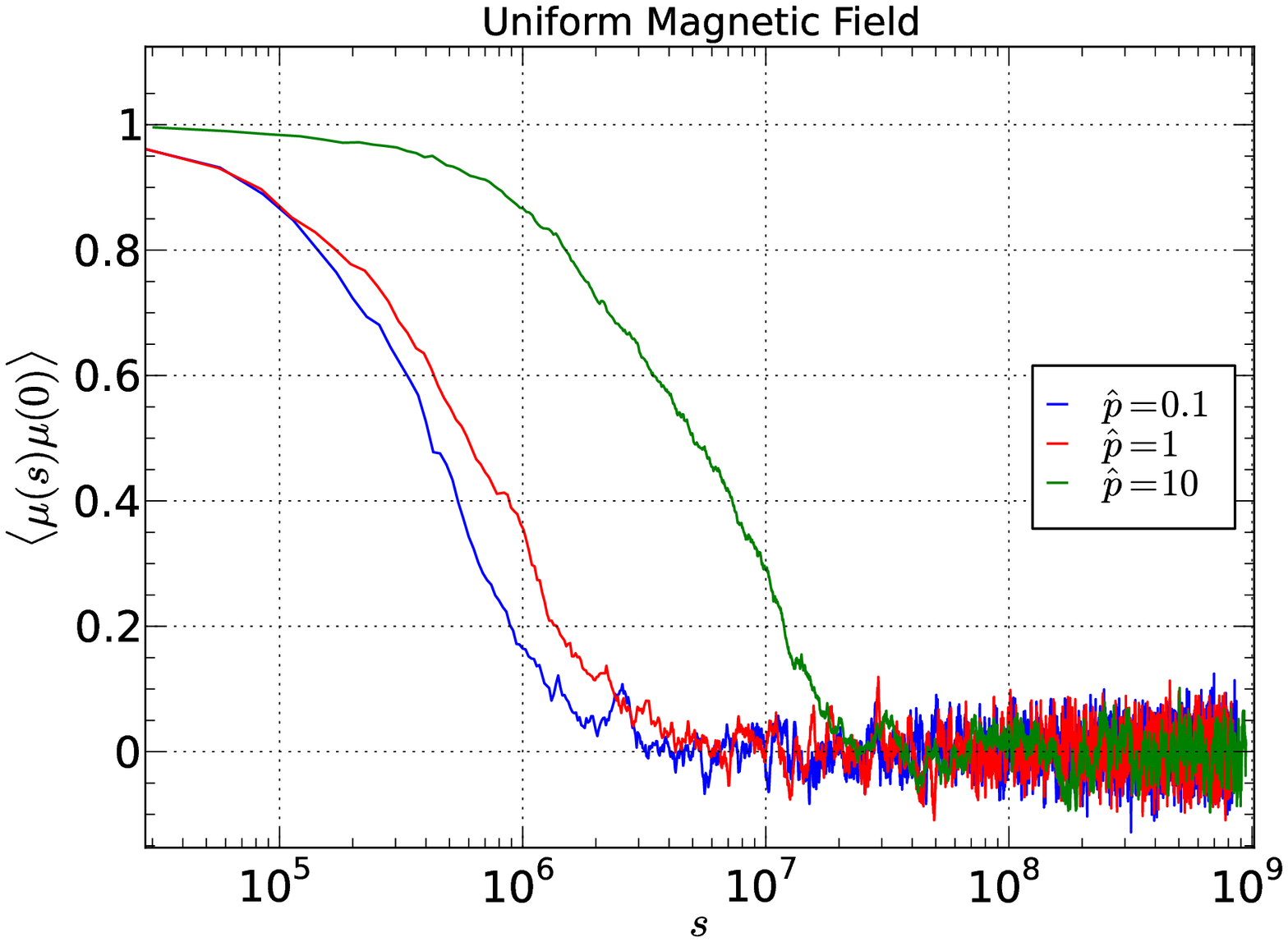}
  }
\subfloat[\label{fig:rundiffuniformd}]{%
  \includegraphics[width=\columnwidth]{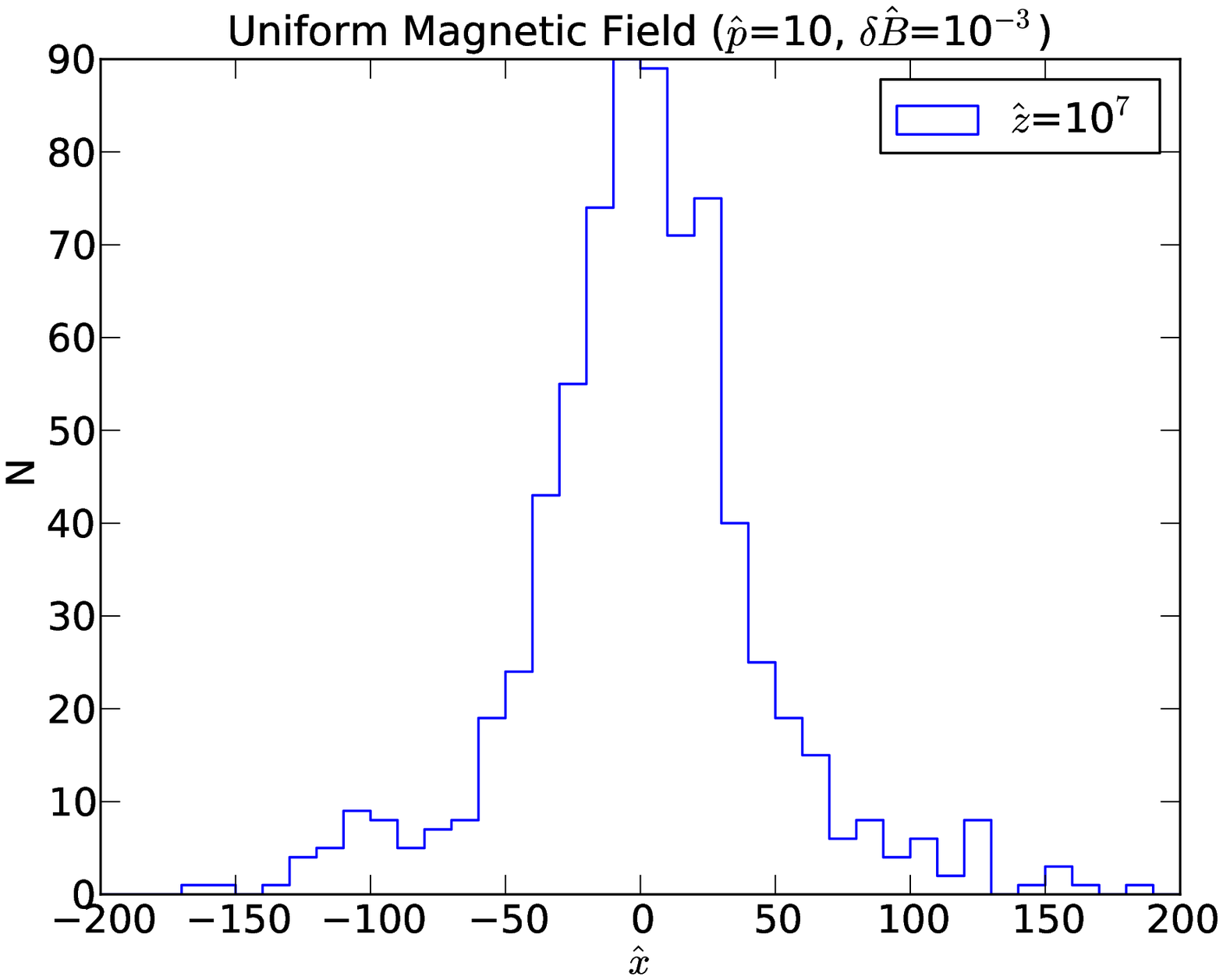}
  }
\caption{Running diffusion coefficient parallel (a) and perpendicular (b) to a uniform magnetic field from numerical particle trajectory integration with pitch angle scattering simulated for $\delta \hat{B}$=10$^{-3}$. The coefficients are determined for an ensemble of 1000 particles with momentum $\hat{p}$ = 0.1, 0.3, 1, 3, 10, 30. It is evident that diffusion regime is reached well within integration time. In panel (c) is the particles pitch angle correlation function for $\hat{p}$ = 0.1, 1, 10. 
In panel (d) is the $\hat x$ spatial coordinate distribution of particles on the $\hat z=10^7$ plane, showing cross fieldline transport.
Particle trajectory integration was calculated for momentum $\hat p$=10, with pitch angle scattering for $\delta \hat B$=10$^{-3}$. The initial spatial distribution at $\hat z=0$ was within $\hat x$=1.}
\label{fig:rundiffuniform}
\end{center}
\end{figure*}

\begin{figure}[!t]
\begin{center}
\subfloat[\label{fig:diffuniforma}]{%
  \includegraphics[width=\columnwidth]{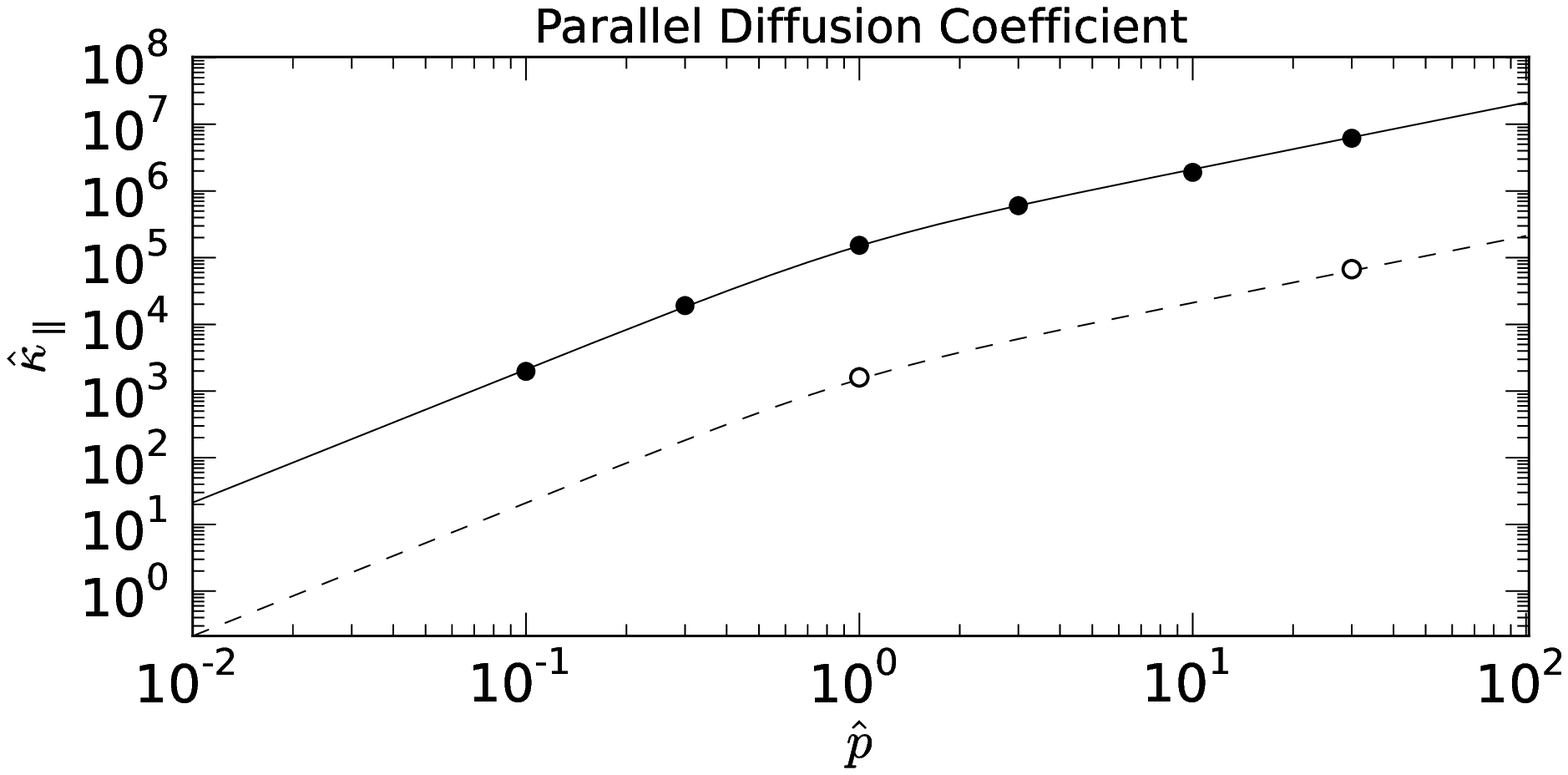}
  }

\subfloat[\label{fig:diffuniformb}]{%
  \includegraphics[width=\columnwidth]{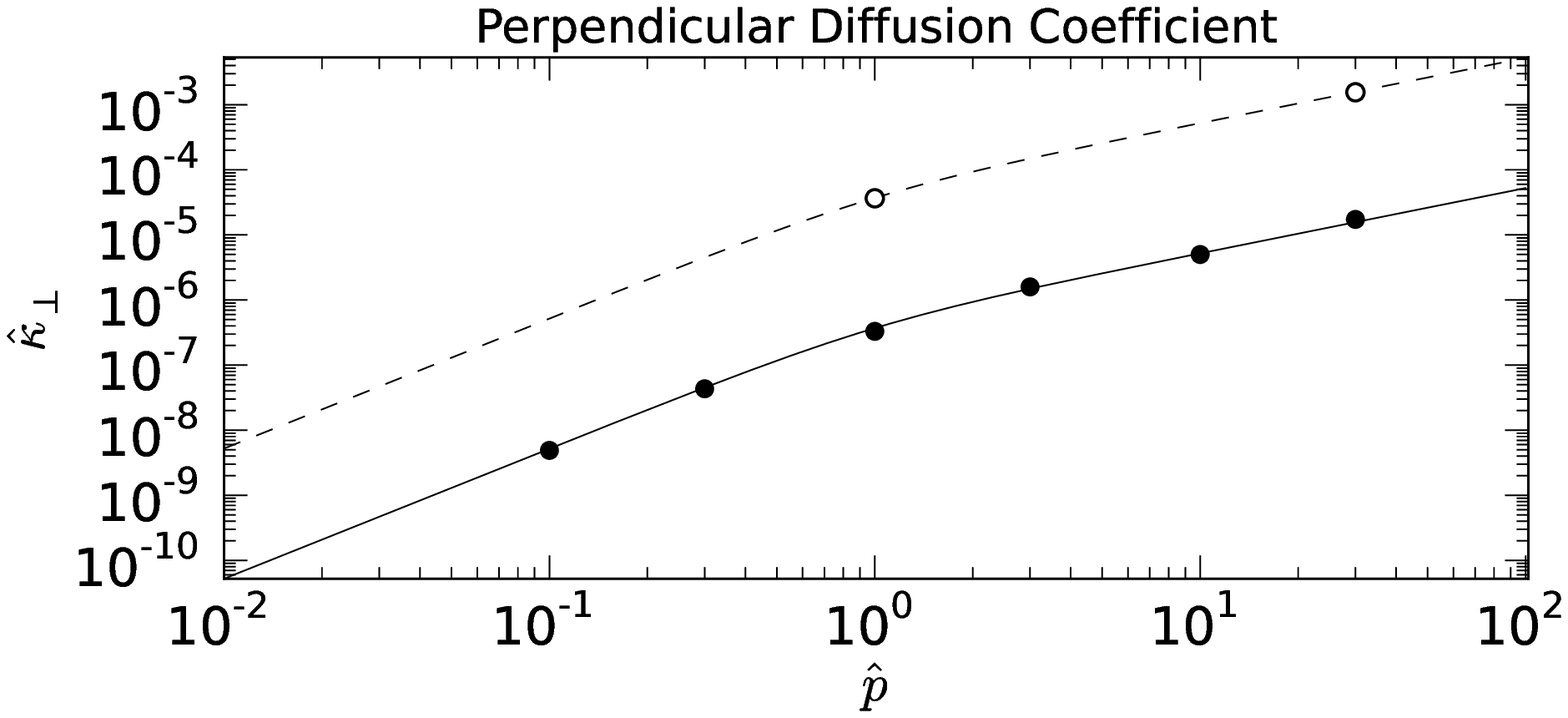}
  }
\caption{Parallel (a) and perpendicular (b) diffusion coefficient in a uniform magnetic field as a function of particle momentum
with pitch angle scattering for $\delta \hat{B}$=10$^{-3}$ (solid circles) and $\delta \hat{B}$=10$^{-2}$ (empty circles) compared with the corresponding expectations from eqs. (\ref{kpar}) and (\ref{kperp}) expressed in non-dimensional coordinates (continuous and dash lines, respectively). Numerical values can be found in Table \ref{tab:uniruns}.}
\label{fig:diffuniform}
\end{center}
\end{figure}

Figure \ref{fig:rundiffuniform} shows how the running parallel (a) and perpendicular (b) diffusion coefficients defined in eqn. (\ref{runningD}) change monotonically and then level off with time. The figures show the effect of pitch angle scattering with $\delta \hat B = 10^{-3}$ on protons with momenta $\hat p$ = 0.1, 0.3, 1, 3, 10, 20. The initial increase of $D_{\parallel}$ is due to ballistic motion of the particles along the fieldlines. On the other hand, $D_{\perp}$ decreases because the perpendicular displacement of the particles is limited by the gyromotion. The running diffusion coefficients settle down to their asymptotic values after about one scattering time $\hat \tau_{sca}=1/\nu$, which according to eqn. (\ref{nu}) depends on $\hat p$ as $\sqrt{1+\hat p^2}$. The transition to diffusion occurs at about the same time that the pitch angle decorrelates, as shown in Figure \ref{fig:rundiffuniformc}.
%
Figure \ref{fig:rundiffuniformd} shows the $\hat x$ coordinate distribution of 1000 on the $\hat z$=10$^7$ plane  (integrated over all times). The particle distribution at initial time was limited in the interval $\hat x <$1. The spread in the particle distribution provides a visualization of the cross fieldline diffusion due to pitch angle scattering. The corresponding parallel and perpendicular diffusion coefficients are shown in Figure~\ref{fig:diffuniform}, for $\delta \hat B$=10$^{-3}$,10$^{-2}$. The numerical calculations are in perfect agreement with the expectations from eqns. (\ref{kpar}) and (\ref{kperp}), that are based on quasilinear theory. We note that estimates of the scattering rates for GeV cosmic rays based on cosmic ray propagation models suggest similarly small
values ($\delta\hat B\sim 10^{-3}$). 

We have found that quasilinear theory breaks down when $\delta \hat B\sim 0.1$.
Since the theory is based on small angle scatterings, while the change  $\Delta\mu$ in the cosine of the pitch angle can be as large as 4.4 $\delta \hat B$, some breakdown in the theory for $\delta \hat B\sim 0.1$ does not surprise us. We do not pursue this topic because it is not the focus of our paper.

%
\begin{figure}[!ht]
\begin{center}
  \includegraphics[width=\columnwidth]{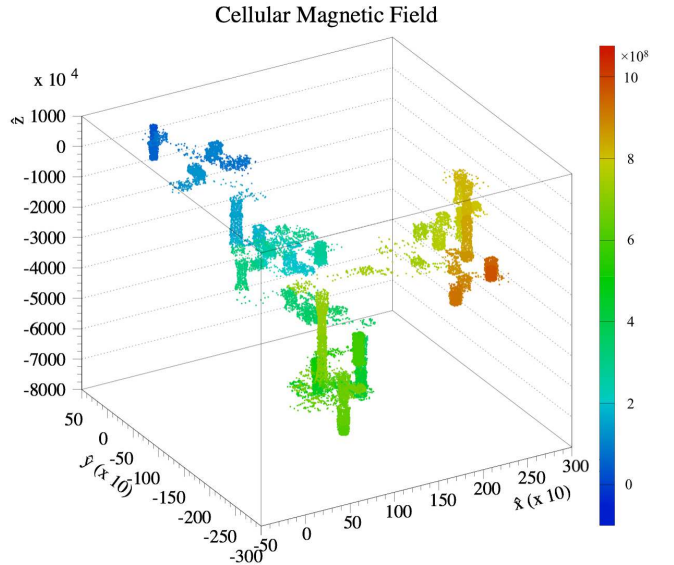}
\caption{The trajectory of a particle with $\hat p=$10 numerically integrated in a cellular magnetic field with $\epsilon=10^{-2}$ in the presence of pitch angle scattering for $\delta \hat B=10^{-3}$. The trajectory is visualized using some of the integration points and the color represents time from $s=0$ (blue) to $s=10^9$ (red).}
\label{fig:celltraj}
\end{center}
\end{figure}
%
%
\subsection{The Cellular Field}\label{ssec:cellular}
\begin{figure*}[!ht]
\begin{center}
\subfloat[\label{fig:rundiffcella}]{%
  \includegraphics[width=0.95\columnwidth]{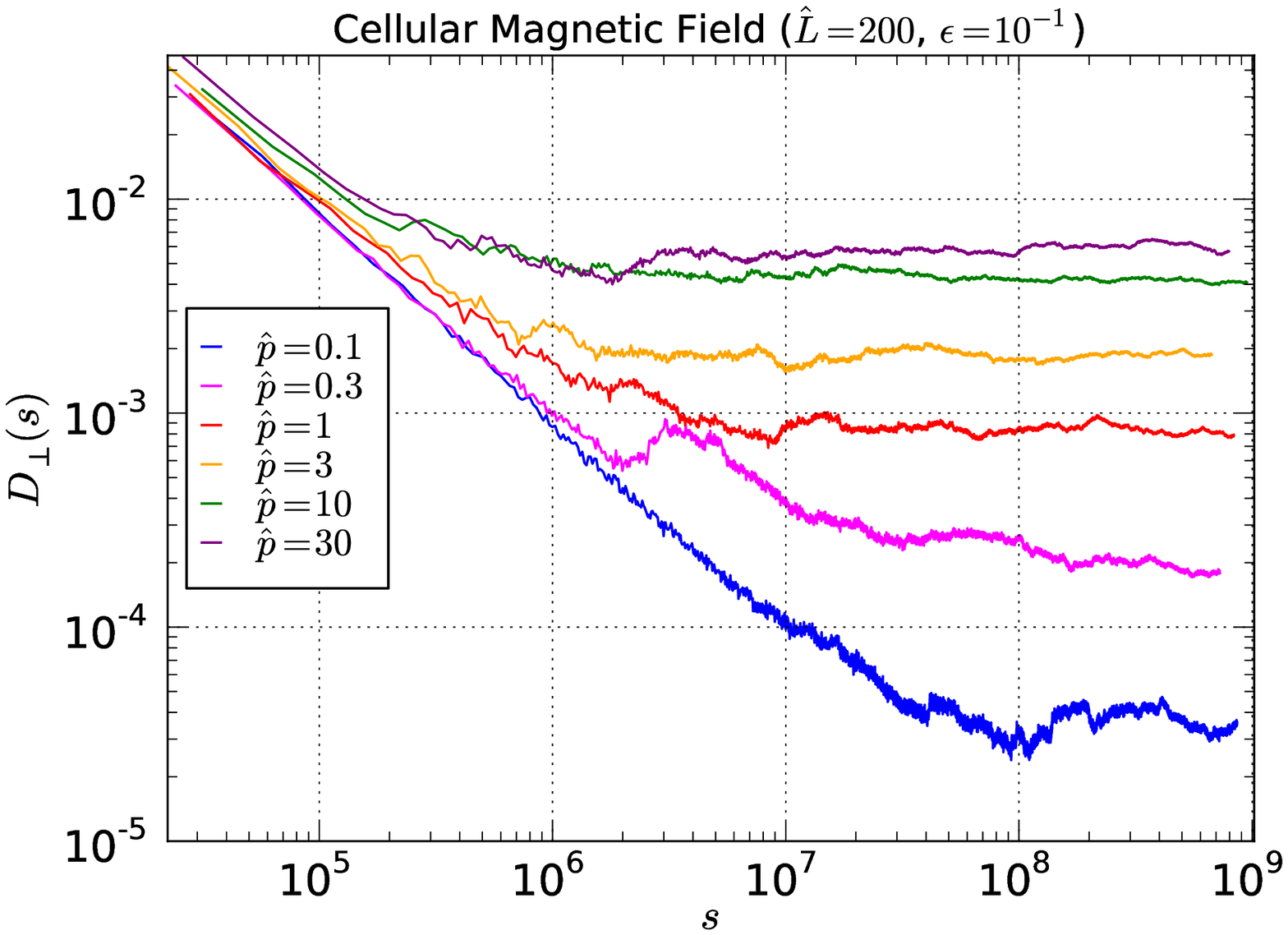}
  }
\subfloat[\label{fig:rundiffcellb}]{%
  \includegraphics[width=0.95\columnwidth]{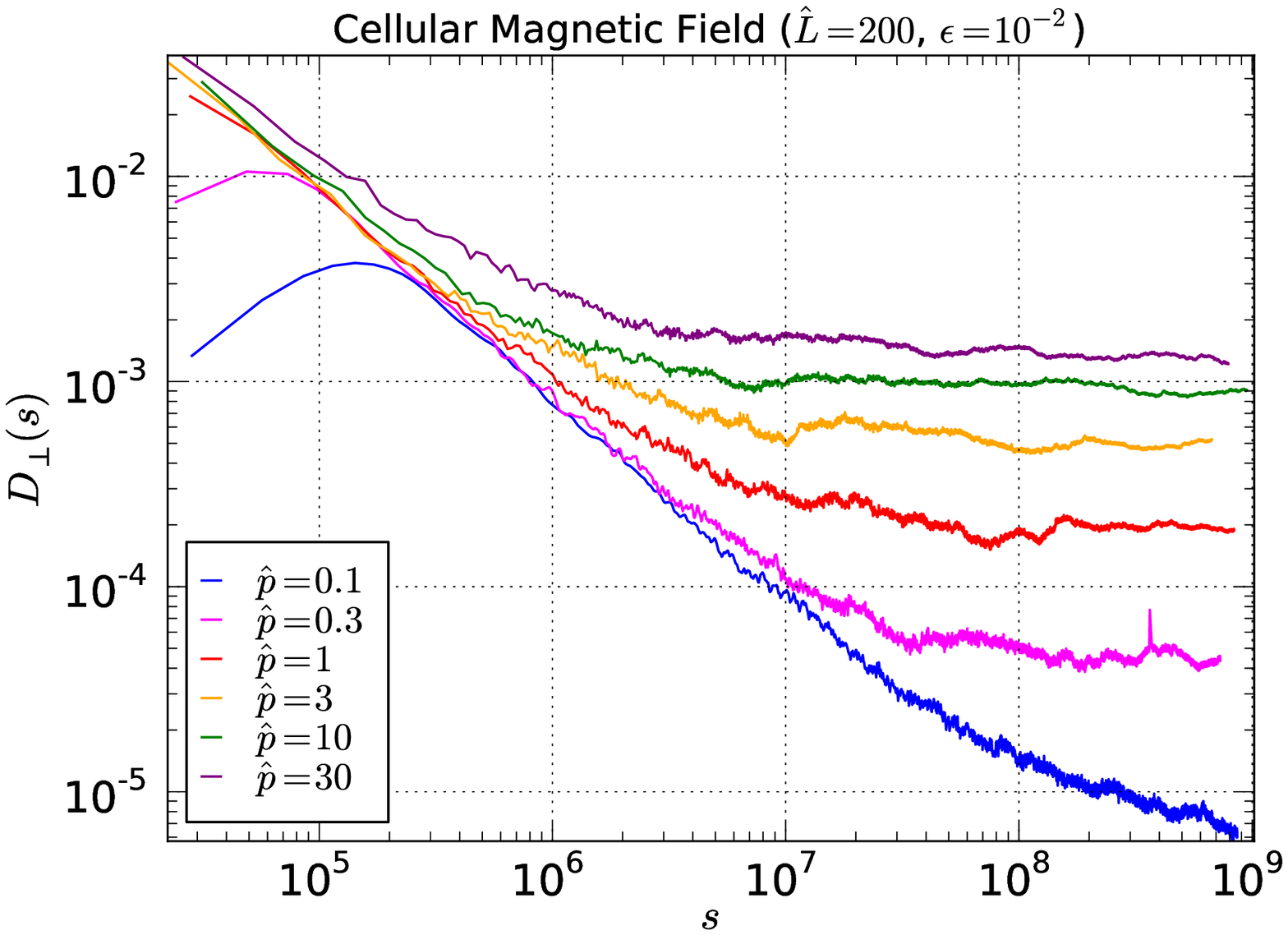}
  }

\subfloat[\label{fig:rundiffcellc}]{%
   \includegraphics[width=0.95\columnwidth]{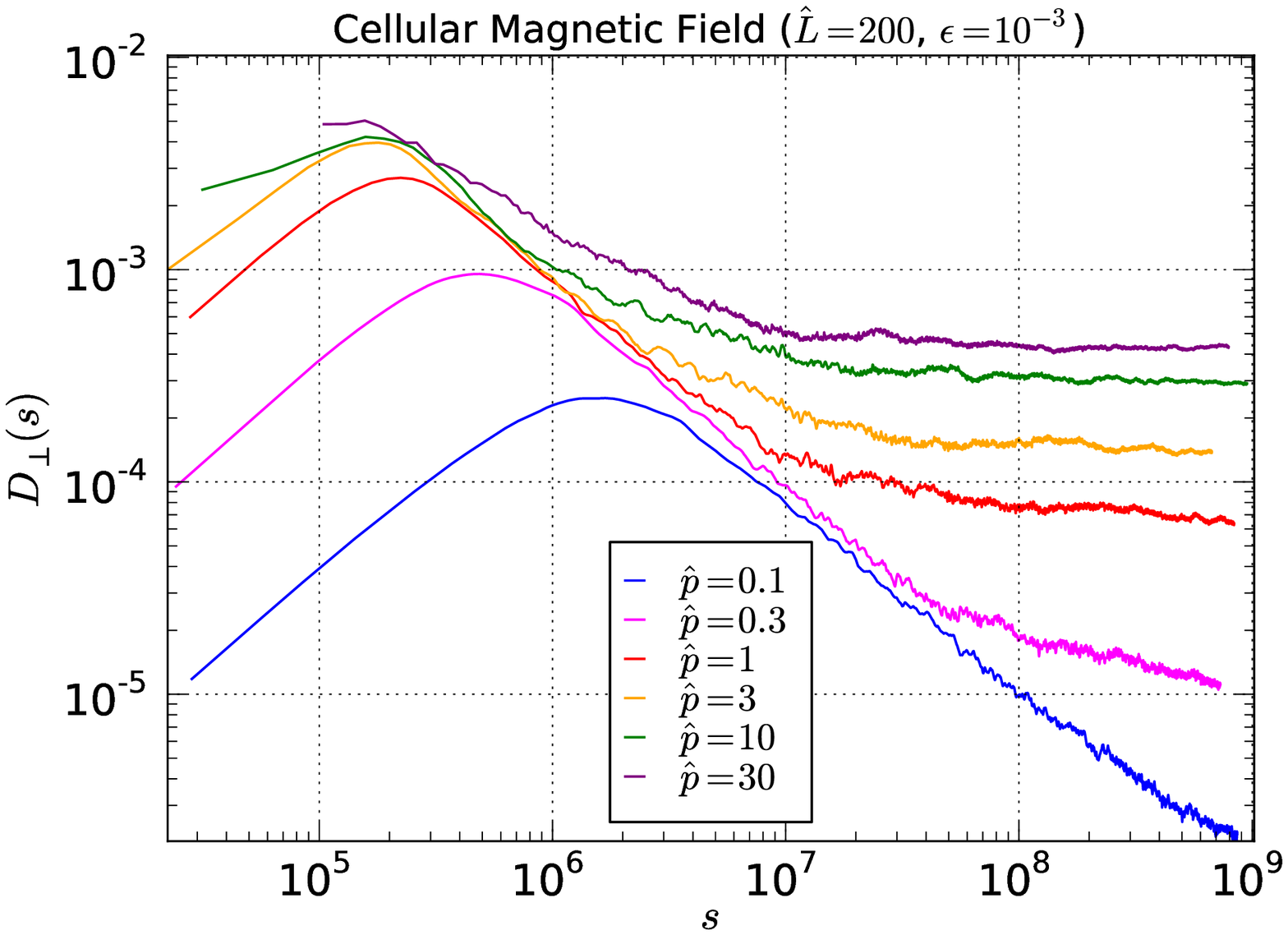}
   }
\subfloat[\label{fig:rundiffcelld}]{%
  \includegraphics[width=0.95\columnwidth]{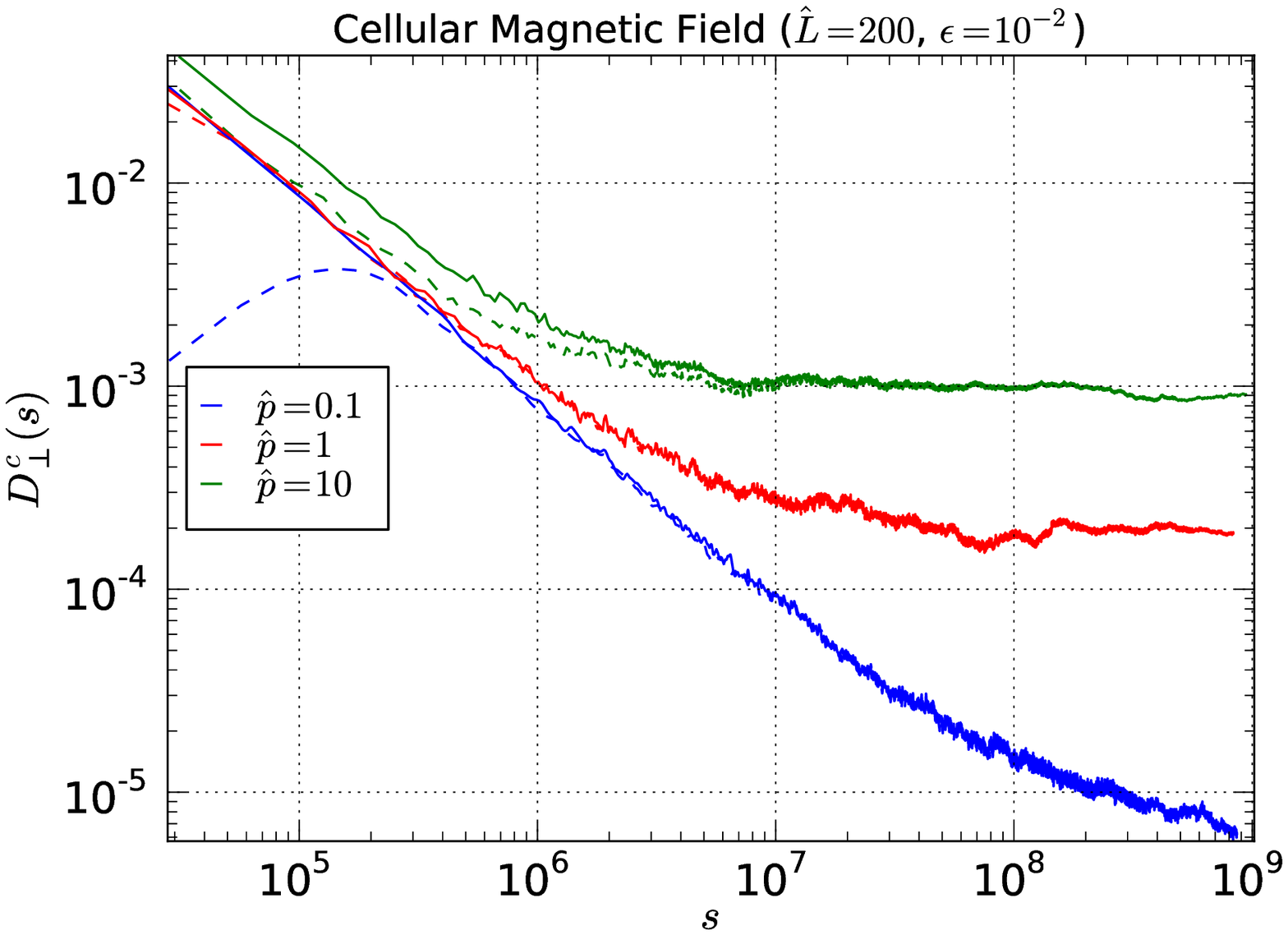}}

\subfloat[\label{fig:rundiffcelle}]{%
   \includegraphics[width=0.95\columnwidth]{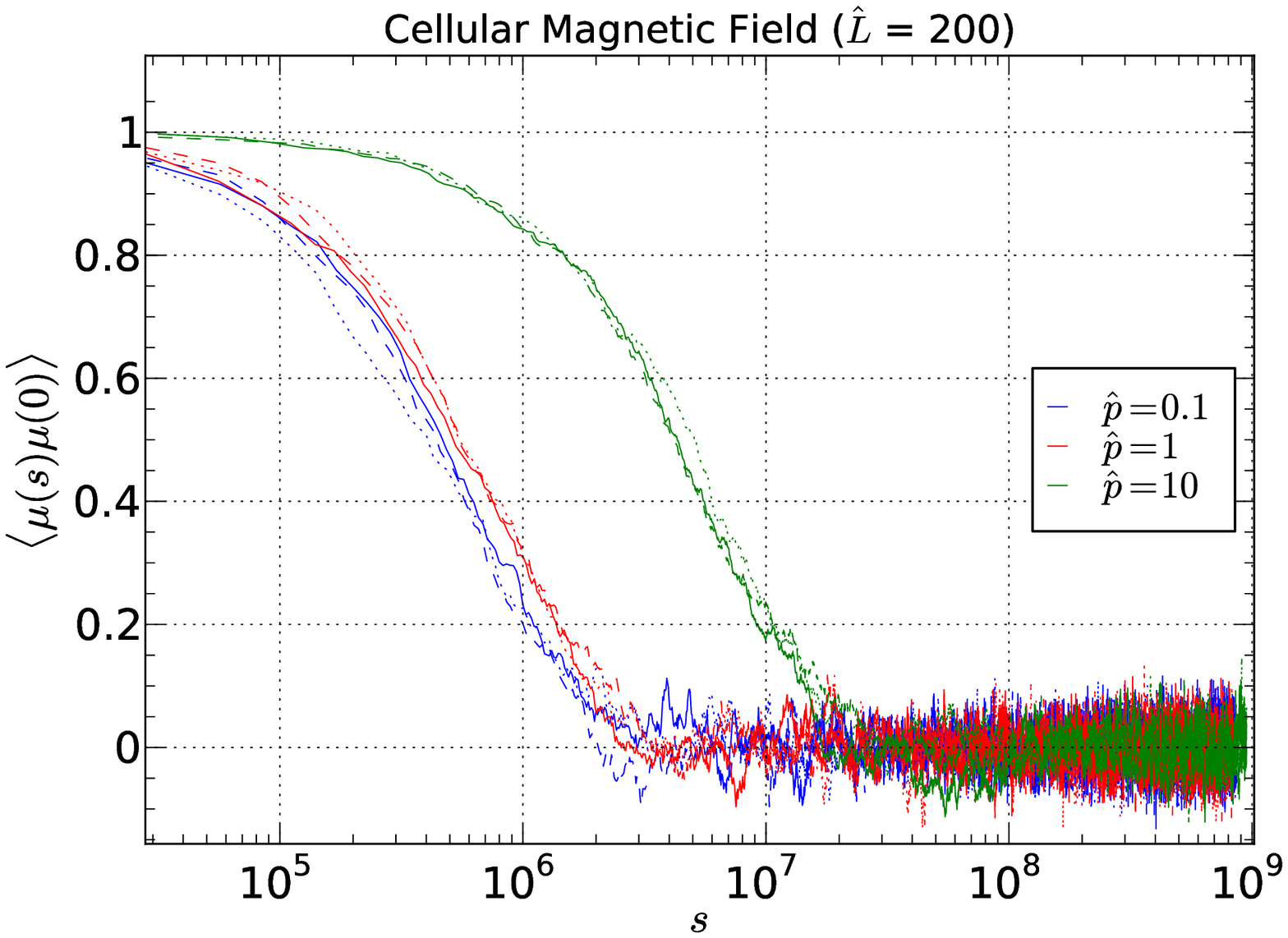}
   }
\subfloat[\label{fig:rundiffcellf}]{%
  \includegraphics[width=0.95\columnwidth]{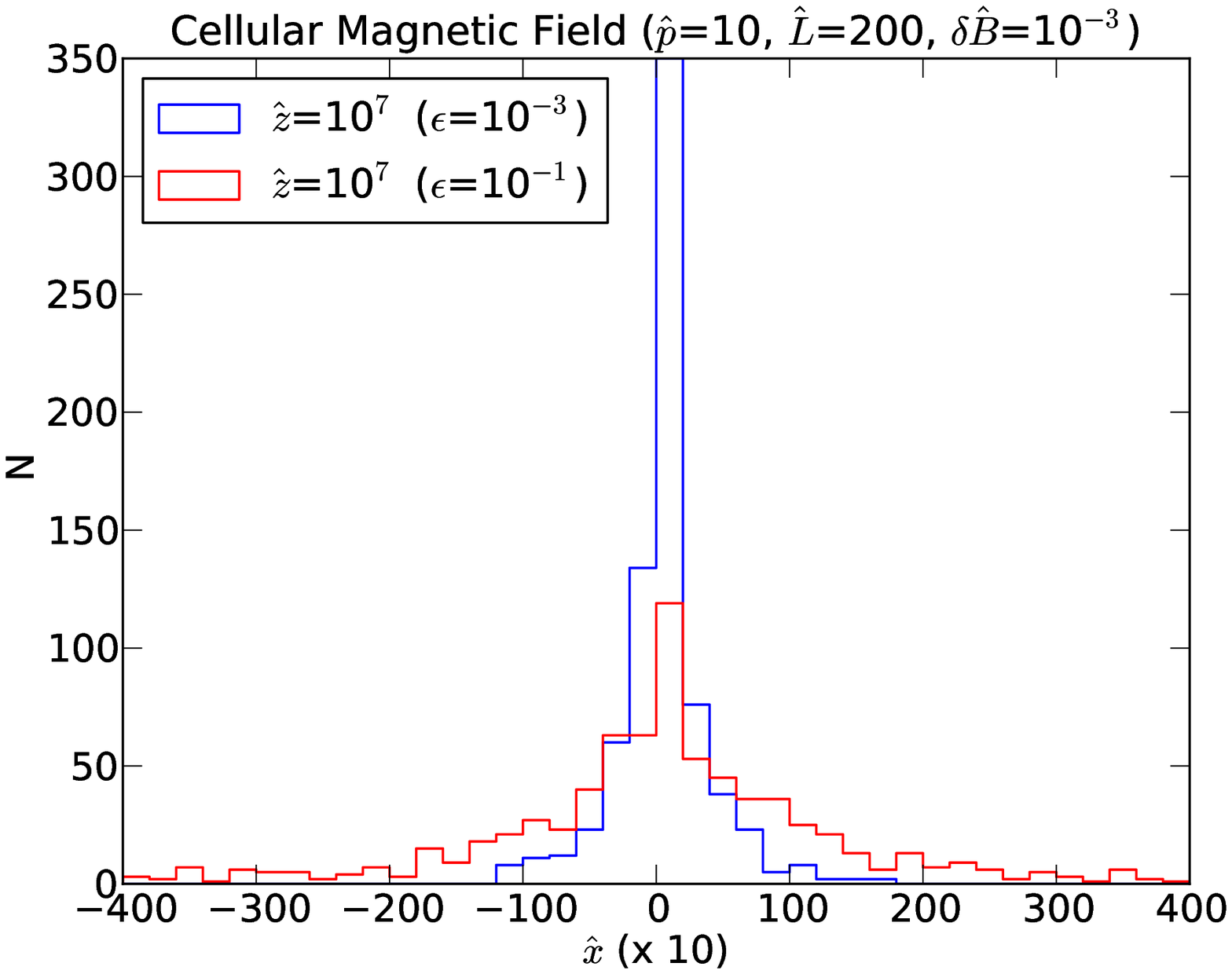}}
\caption{Running diffusion coefficient perpendicular to the guide field in the cellular magnetic geometry with $\epsilon$ = 10$^{-1}$ (a), 10$^{-2}$ (b) and 10$^{-3}$ (c) obtained with the numerical particle trajectory integration and pitch angle scattering simulated with magnetic perturbation strength $\delta \hat{B}$=10$^{-3}$. The coefficients are determined for an ensemble of 1000 particles with momentum $\hat{p}$ = 0.1, 0.3, 1, 3, 10, 30. Note the statistical fluctuations affecting the case with relatively large magnetic perturbation relative strength. In panel (d) the cross-fieldline running perpendicular diffusion coefficient converges to the same value of that calculated with respect to the guide field (see \S\ref{subsec:spatialdiffusion}). In panel (e) the particles pitch angle correlation function for $\hat{p}$ = 0.1, 1, 10, and for $\epsilon$ = 10$^{-1}$, 10$^{-2}$, 10$^{-3}$ (continuous, dashed and dotted lines, respectively). 
In panel (f) is the $\hat x$ spatial coordinate distribution of particles on the $\hat z=10^7$ plane, for $\hat p$=10 and $\delta \hat B$=10$^{-3}$. The initial spatial distribution at $\hat z=0$ was within $\hat x$=50.
}
\label{fig:rundiffcell}
\end{center}
\end{figure*}

As shown in \S\ref{ssec:uniform} particles propagating in a uniform magnetic field become diffusive on a time scale $\hat \tau_{sca}$ in the presence of pitch angle scattering.

With our numerical integration of particle trajectories we show that cross fieldline motion is significantly enhanced if large scale structures in the magnetic fields exist. Even with simple perpendicular geometrical structures, we find a breakdown of quasilinear theory for $\delta \hat B\ll 1$.

For each of the structured magnetic field configurations described in \S\ref{ssec:bfields}, eqn. (\ref{dimlessmomentum}) is solved for 1000 protons with initial isotropic direction distribution and with momenta $\hat p = 0.1, 0.3, 1, 3, 10, 30$. The particle gyroradii are always smaller than $\hat L/2 = 100$, which is the cell size for both the cellular and GP configurations. Each particle trajectory is integrated in an unbounded spatial volume up to a maximum time of $s_f=10^9$. 
Energy losses are not taken into account in the trajectory integration, therefore particle energy is conserved within the limits of numerical accuracy (see Appendix). Trajectories are integrated both without and with pitch angle scattering, treated as a random process. The fiducial magnetic perturbation strength used to simulate scattering is $\delta \hat B$ = 10$^{-3}$. In some cases the values $\delta \hat B$ = 10$^{-2}$ - 10$^{-1}$ are used to study the dependence on scattering frequency.

%
Particles propagating in such cellular magnetic field geometry remain confined within a given cell
unless something causes cross-field motion to the nearby cells, as shown in Figure \ref{fig:celltraj}. Curvature and gradient drifts are too small to cause particles to move to other cells
unless the particles are within a gyroradius of a cell boundary. On the other hand, 
pitch angle scattering reduces
confinement to a given cell causing particles to move across fieldlines.

The running diffusion coefficients in the presence of pitch angle scattering are shown in Figure~\ref{fig:rundiffcell} for three different values of the transverse fieldstrength parameter $\epsilon$=10$^{-3}$, 10$^{-2}$, 10$^{-1}$. Generally, each $D_{\perp}$ displays three regimes of behavior. Initially, $D_{\perp}$ 
increases with time, particularly for the smaller values of $\epsilon$ and lower particle energies. This can be understood as follows. The maximal orbital time of a particle around one magnetic cell is given by $\tau_{orb}\approx l / \langle v_{\parallel} \rangle \epsilon$, with $l$ the cell perimeter and $\langle v_{\parallel} \rangle \epsilon = \frac{v}{2}\epsilon$ the component of parallel particle velocity perpendicular to the guide magnetic field averaged over particle pitch angle. From eqn. (\ref{cellularA}), $l = 2\hat{L}$. In dimensionless coordinates, the orbital time $\tau_{orb}$ is
\begin{equation}\label{eq:torb}
\hat{\tau}_{orb}\approx \frac{4\hat{L}}{\epsilon}\frac{\sqrt{1+\hat{p}^2}}{\hat{p}}.
\end{equation}
Values of $\hat \tau_{orb}$ are given in the Appendix tables.
%
%
%
%
\begin{figure}[!ht]
\begin{center}
  \includegraphics[width=\columnwidth]{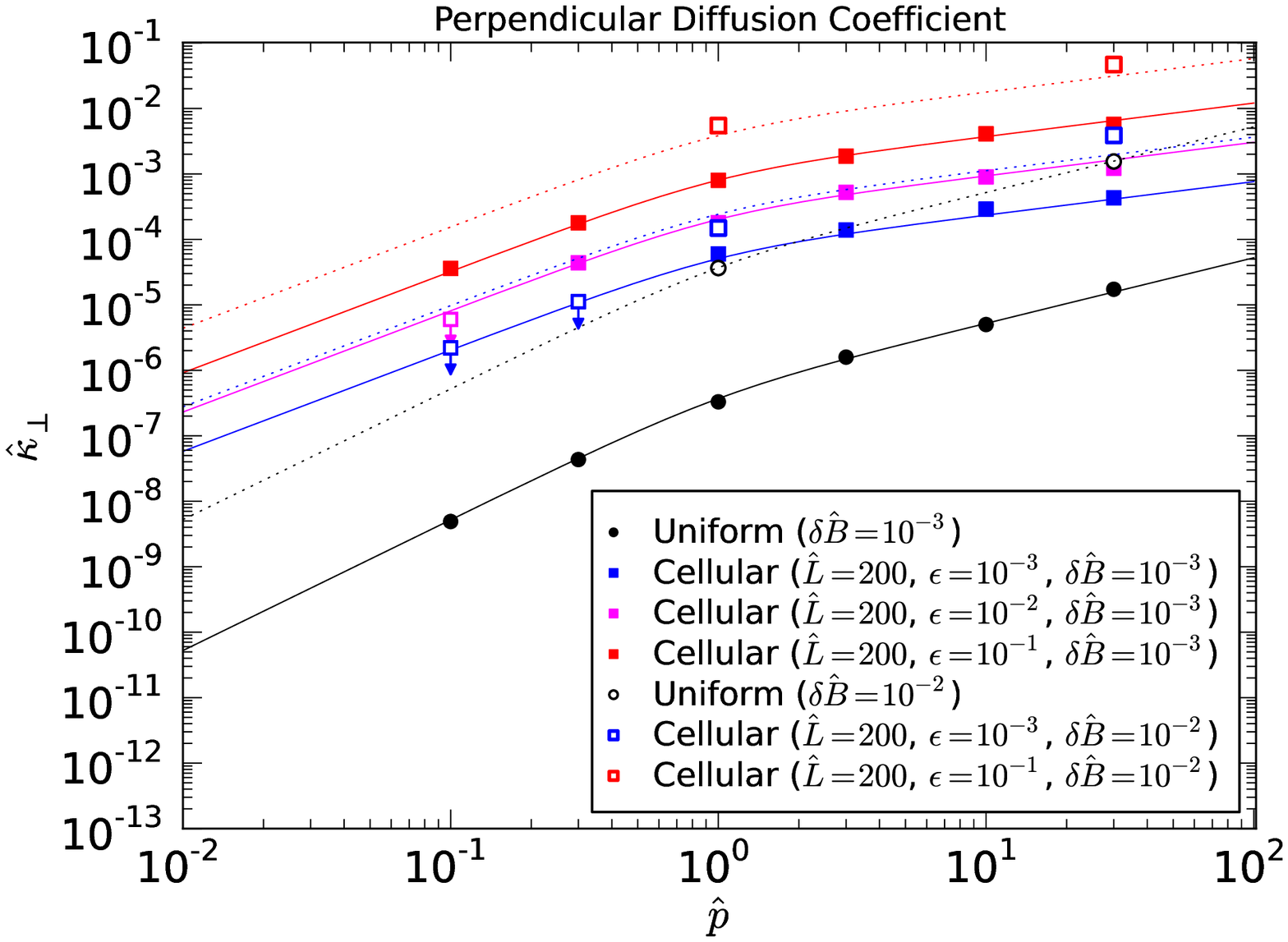}
\caption{Perpendicular diffusion coefficient in the cellular magnetic geometry for magnetic perturbation $\delta \hat B$ = 10$^{-3}$ (full symbols) and $\delta \hat B$ = 10$^{-2}$ (empty symbols), compared to the corresponding coefficients in the uniform magnetic field. The empty symbols with the downward arrow indicate the cases where diffusion regime was not reached within the integration time (as shown in Figure~\ref{fig:rundiffcell}). The black lines are the quasilinear theory prediction eqn. (\ref{kperp}), and the colored lines correspond to the fit result eqn. (\ref{eq:cellfit}). Numerical values can be found in Table \ref{tab:cellruns}.}
\label{fig:diffcell}
\end{center}
\end{figure}
%
%
The orbital motion around a magnetic cell contributes to perpendicular displacement with respect to the guide magnetic field up to a time scale $\hat \tau_{orb}/2$, when particles feel the confinement within the cell. This explains the initial ballistic behavior of $D_{\perp}$. As discussed in \S\ref{subsec:spatialdiffusion}, this initial perpendicular displacement does not necessarily represent cross-field transport, due to the fieldlines helical shape around the cells.
However, the maximum fieldline displacement is $\sqrt{200}$. Thus, to the extent that the particles are tied to the fieldlines, $D_{\perp}$ starts to decrease once the particles have completed about one orbit. This is the second stage. Figure \ref{fig:rundiffcelld} shows the cross-fieldline running perpendicular diffusion coefficient, described in \S\ref{subsec:spatialdiffusion}, compared to that with respect to the guide field. In the cross-fieldline case the ballistic behavior is obviously absent, although the two coefficients converge to the same value when they approach diffusion regime.

In order to reach the diffusion regime, which is the third stage, particles must be able to migrate to nearby cells.
As discussed earlier, particle decorrelation occurs at scattering time $\hat \tau_{sca}$, regardless of the large scale magnetic field structure, since it is determined by the gyro scale magnetic perturbations, effectively accounted by the simulation of pitch angle scattering. Figure ~\ref{fig:rundiffcelle} shows that the pitch angle correlation functions for different amplitudes of the cellular magnetic field structure, are very similar to that in uniform magnetic field. However, contrary to the case with no large scale magnetic structure, the diffusion regime does not necessarily turn on at the scattering time. Particles with small gyroradius remain confined within a cell even after they are decorrelated. The larger the gyroradius, the easier is for particles to migrate to the nearby cells. An individual particle may go through a sequence of trapping within a cell and of escaping to another one as a consequence of pitch angle scattering, as shown in Figure \ref{fig:celltraj}. Escape from a cell implies transverse motion of
order the cell size.
Figure \ref{fig:rundiffcellf} shows the $\hat x$ coordinate distribution of 1000 particles in the cellular magnetic field, on the $\hat z$=10$^7$ plane  (integrated over all times). The particle distribution at initial time was limited in the interval $\hat x<$50. The spread in particle distribution is larger than in the uniform magnetic field case (Figure \ref{fig:rundiffuniformd}) and it increases with with the amplitude $\epsilon$ of the cellular magnetic field structure.

We found that stable diffusion coefficients are reached when, after a sufficiently long time, the trapping and escaping events are equilibrated. Such coefficient is shown in Figure~\ref{fig:diffcell} as a function of particle momentum for different values of $\epsilon$ and of pitch angle scattering rates ($\delta \hat B$ = 10$^{-3}$ - 10$^{-2}$).
Compared to the case with no magnetic structure, cellular geometry significantly enhances perpendicular diffusion as long as the particle gyroradius is smaller than the cell spatial scale. We find that particles with gyroradii much larger than the cell size are relatively unaffected affected by the cells. However, particles with $\vert\mu\vert$ sufficiently close to 1 can have gyroradii comparable or less than the cell size, and their diffusion is enhanced. Therefore, $\hat \kappa_{\perp}$ for an isotropic distribution of particles in a cellular field is always enhanced over the value for a uniform field.

%
%
\begin{table}[!ht]
\caption{\label{tab:cellfit} Fit parameters for\\Cellular magnetic structure}
\centering
\begin{tabular}{ c r }
\hline
parameter        & value($\pm$ error) \\
\hline
$\ln(C)$            & -4.54 $\pm$ 0.63 \\
$a_L$               & 0.71 $\pm$ 0.11 \\
$a_{\epsilon}$  & 0.59 $\pm$ 0.03 \\
$a_B$              & 0.68 $\pm$ 0.05 \\
$a_p$              & 1.56 $\pm$ 0.11 \\
$a_{p2}$         & -0.53 $\pm$ 0.08 \\
\hline
\end{tabular}
\end{table}
%

In order to quantify the scaling of transverse diffusion coefficients for the cases analyzed in this work, we have calculated trajectories by varying other input parameters such as cell size $\hat L$ and pitch angle scattering frequency, via $\delta \hat B$ for the cellular case, in addition to the mentioned datasets (see Appendix). In the cellular magnetic structure case we use the transverse diffusion coefficients for all the generated datasets and perform a multi-dimensional least squares fit to the function ln$\hat \kappa^C_{\perp}$ where
\begin{equation}
\hat \kappa^C_{\perp} = C\, L^{a_L}\, \epsilon^{a_{\epsilon}}\, \delta \hat B^{a_B}\, \hat p^{a_p}\, (1+\hat p^2)^{a_{p2}},
\label{eq:cellfit}
\end{equation}
where the fit results are summarized in Table~\ref{tab:cellfit}, and shown in Figure~\ref{fig:diffcell}. The errors are estimated using the covariance matrix. It is worth to note that particle trajectories calculated for larger cell sizes did not reach the diffusion regime in the time allowed for the calculation, therefore the fit value for $a_L$ is to be considered an upper bound. We found that omitting $\hat L$ in our fit does not modify the dependence on the other parameters.

From the table we see that we can express the diffusion coefficient as $\hat \kappa_{\perp}^C\approx \hat L^{0.7}\, \delta \hat B^{0.7}\, (\epsilon\, \hat p)^{0.6}\, v$.

\subsection{The Galloway-Proctor Field}\label{ssec:GP}
\begin{figure}[!ht]
\begin{center}
  \includegraphics[width=\columnwidth]{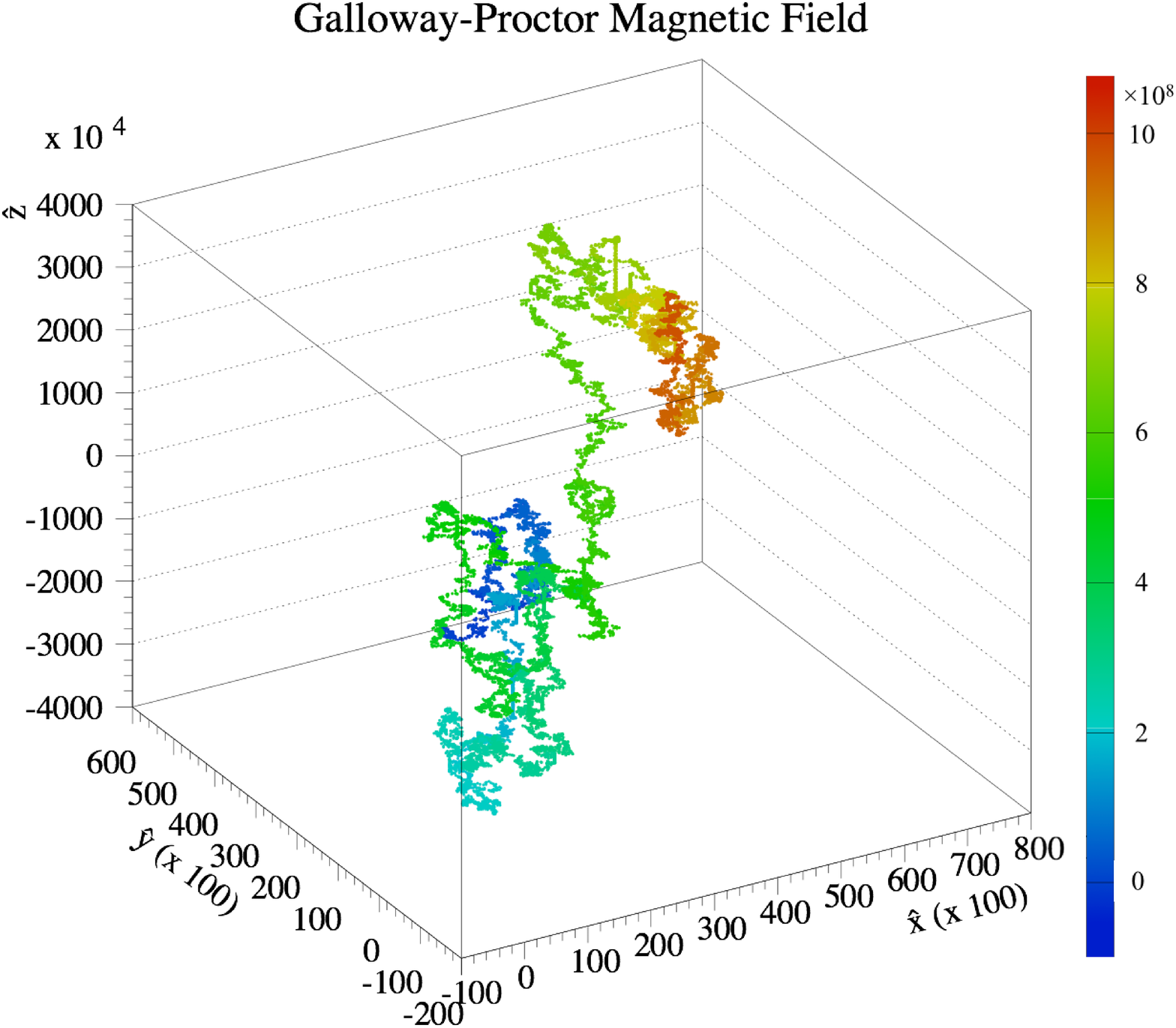}
\caption{Similar to Figure \ref{fig:celltraj} but for a GP magnetic field structure with $\epsilon=10^{-1}$, $\hat \lambda = 0.25$ and $\hat \tau \approx \hat \tau_{orb}$.}
\label{fig:gptraj}
\end{center}
\end{figure}
\begin{figure}[!ht]
\begin{center}
\subfloat[\label{fig:rundiffcellLgpa}]{%
  \includegraphics[width=\columnwidth]{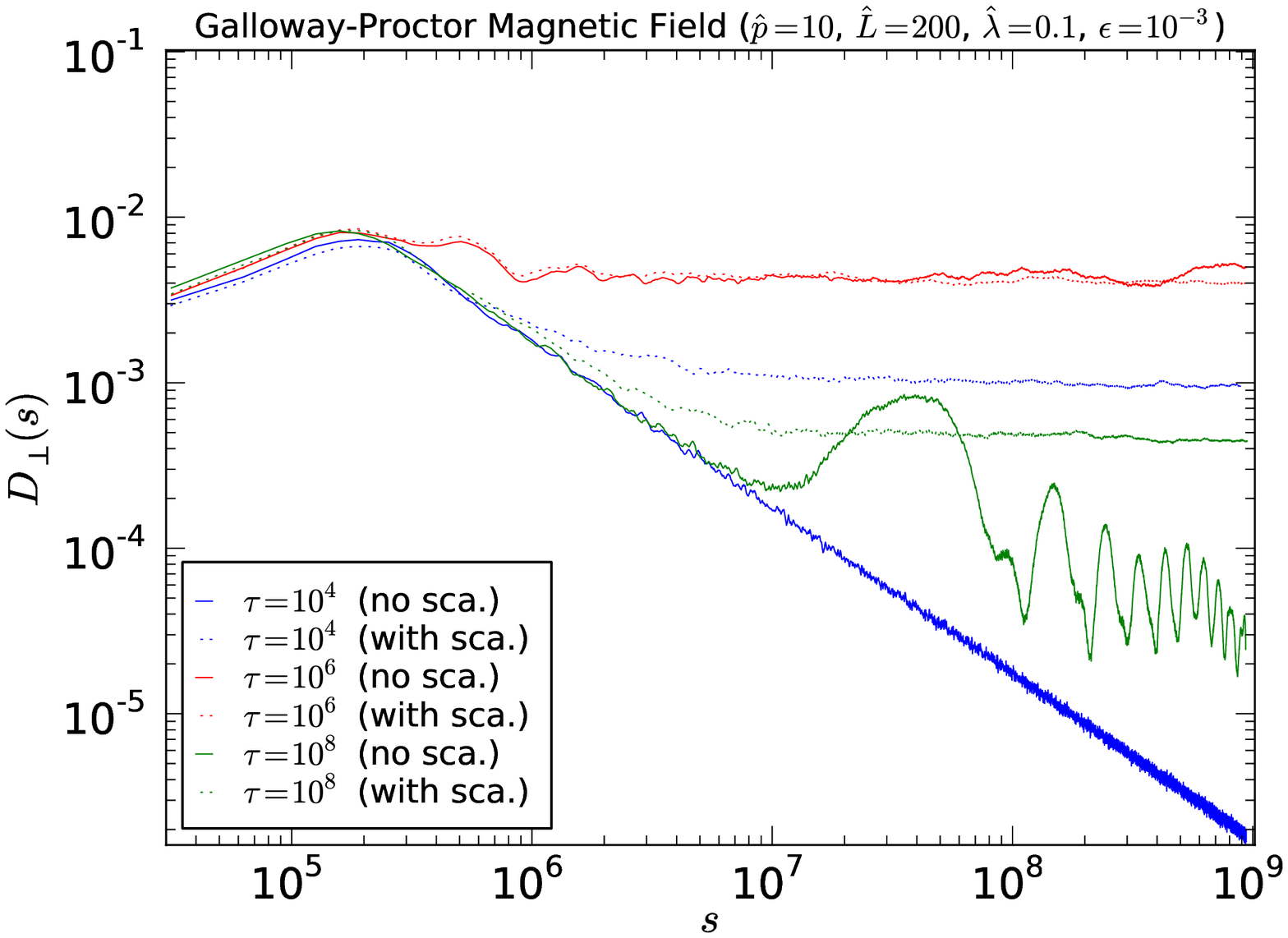}
  }

\subfloat[\label{fig:rundiffcellLgpb}]{%
  \includegraphics[width=\columnwidth]{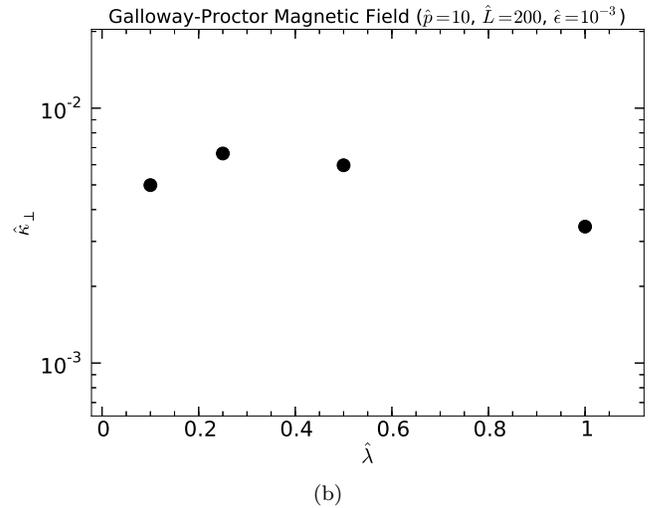}
  }
\caption{Running diffusion coefficient perpendicular to the guide field in the GP magnetic geometry (a) with $\epsilon$ = 10$^{-3}$, $\lambda$ = 0.1 and for different values of $\hat \tau$, compared to orbital time around a magnetic cell of $\hat \tau_{orb}\approx 8\times 10^5$. The coefficient was obtained with the numerical trajectory integration of 1000 particles with momentum $\hat{p}$ = 10 without and pitch angle scattering simulated with magnetic perturbation strength $\delta \hat{B}$=10$^{-3}$. Diffusion coefficient (b) for $\hat p$ = 10 and $\epsilon$ = 10$^{-3}$ for different values of $\hat \lambda$ in the resonance condition $\hat \tau_{sca}\approx \hat \tau_{orb}$. Numerical values can be found in Table \ref{tab:gpruns}.}
\label{fig:rundiffcellLgp}
\end{center}
\end{figure}
\begin{figure*}[!ht]
\begin{center}
\subfloat[\label{fig:rundiffgp2a}]{%
  \includegraphics[width=\columnwidth]{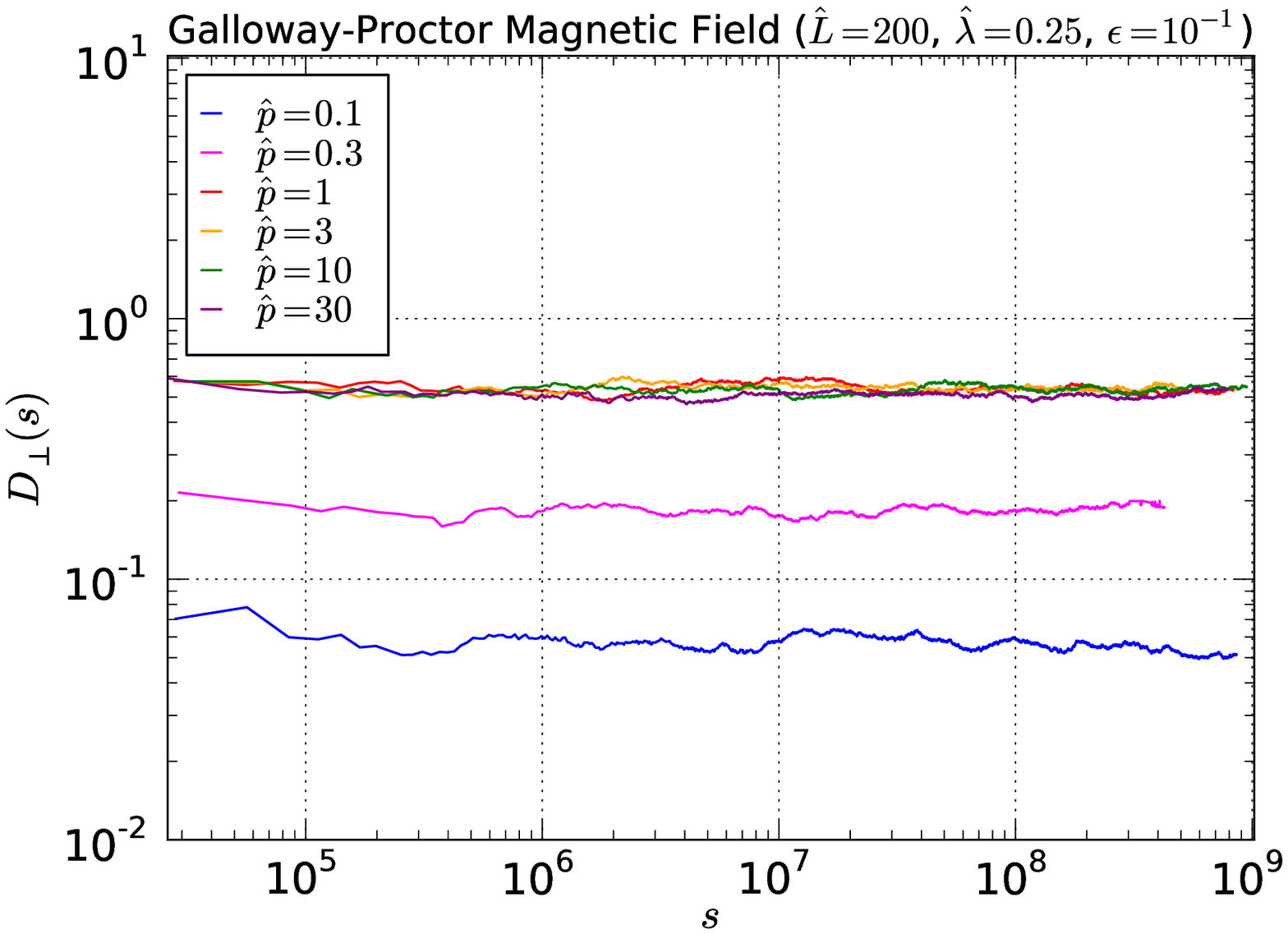}
  }
\subfloat[\label{fig:rundiffgp2b}]{%
  \includegraphics[width=\columnwidth]{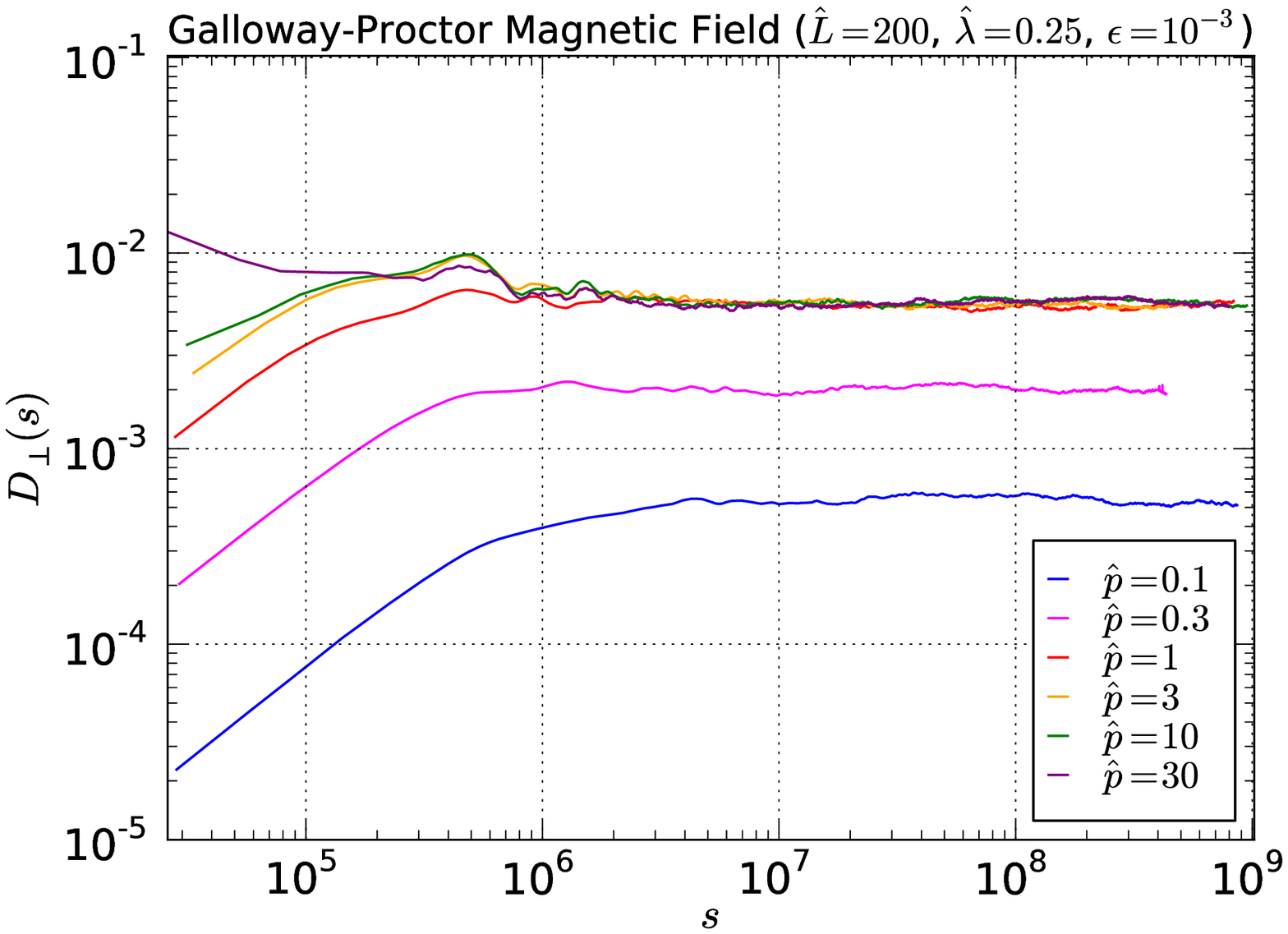}
  }

\subfloat[\label{fig:rundiffgp2c}]{%
  \includegraphics[width=\columnwidth]{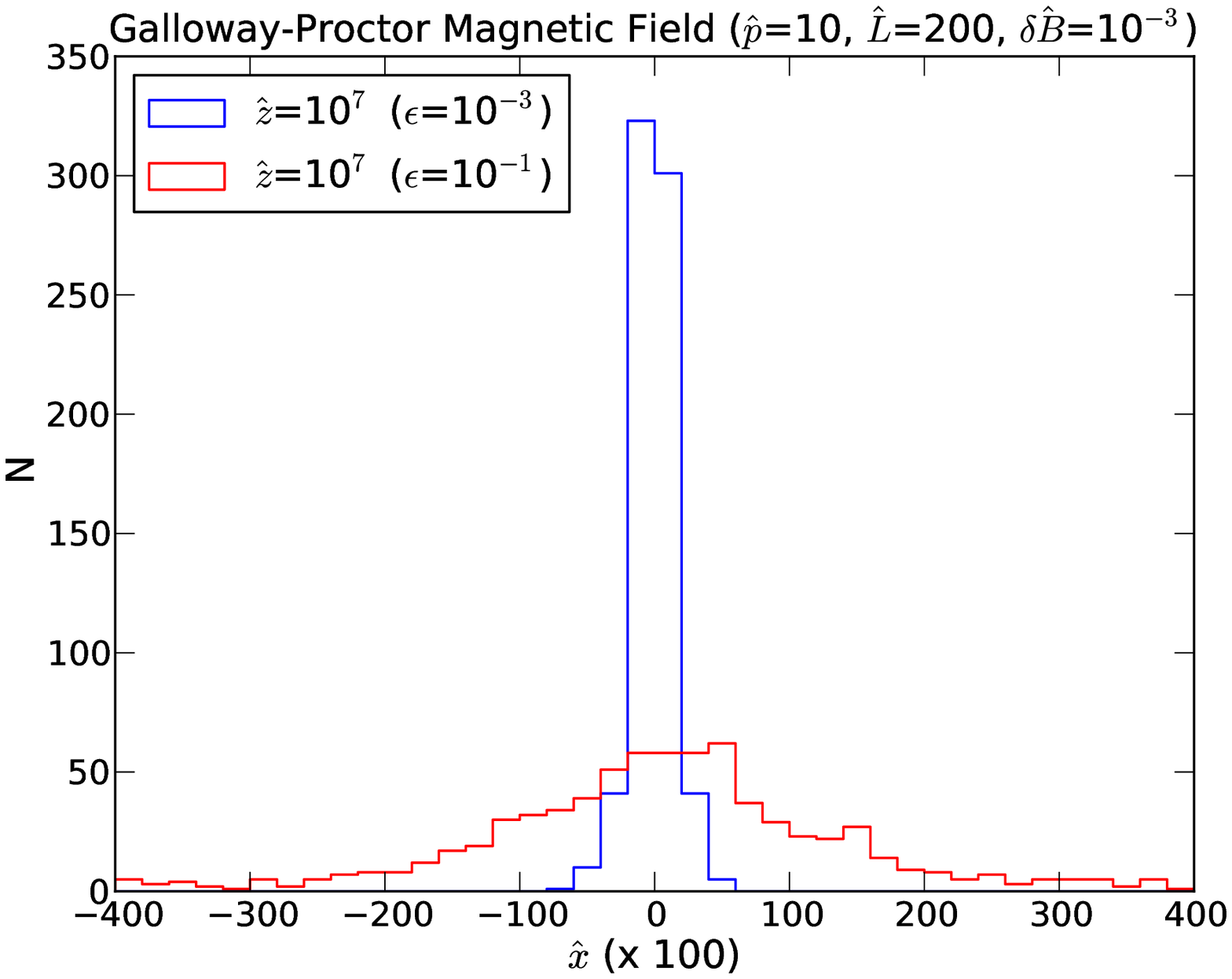}
  }
\caption{Running diffusion coefficient perpendicular to the guide field in the GP magnetic geometry with $\epsilon$ = 10$^{-1}$ (a) and 10$^{-3}$ (b), $\lambda$ = 0.25 and $\hat \tau \approx \tau_{orb}$. The coefficient was obtained with the numerical trajectory integration of 1000 particles with momentum $\hat{p}$ = 0.1, 0.3, 1, 3, 10, 30, and pitch angle scattering simulated with magnetic perturbation strength $\delta \hat{B}$=10$^{-3}$. 
In panel (c) is the $\hat x$ spatial coordinate distribution of particles on the $\hat z=10^7$ plane, for $\hat p$=10 and $\delta \hat B$=10$^{-3}$. The initial spatial distribution at $\hat z=0$ was within $\hat x$=50.}
\label{fig:rundiffgp2}
\end{center}
\end{figure*}
The GP transverse magnetic field structure is more complicated because of the time dependence of the magnetic cell position. 
By analogy with the GP {\it flow}, the magnetic streamlines are known to be chaotic. The
 average Lyapunov exponent, which characterizes the separation rate of infinitesimally close points in the flow,  has a maximum for the so-called complexity parameter $\hat \lambda \approx 0.25$
\citep{brummell_2001}. As mentioned in \S 2.1, if the magnetic flux function $A_{GP}$ is interpreted as a velocity stream function, the resulting flow is known to have good mixing properties, and to quickly disperse passive scalar particles \citep{heitsch_2004}. It is these good mixing properties of the GP {\textit{flow}} that motivated our  study of diffusion in the GP magnetic field.

To some extent, these mixing properties carry over to particles propagating on magnetic fieldlines, as shown in Figure \ref{fig:gptraj}. Numerical integration of particle trajectories shows that even with no pitch angle scattering there may be diffusive behavior, depending on the cell turnover time $\hat \tau = \frac{2\pi}{\hat \omega}$ and how it compares with particle orbital time $\hat \tau_{orb}$. Figure~\ref{fig:rundiffcellLgpa} shows the running perpendicular diffusion coefficient for $\hat{p}$ = 10, $\hat{L}$ = 200 and $\epsilon$ = 10$^{-3}$, in the case of $\hat{\lambda}$ = 0.1 for three different values of $\hat \tau$. If $\hat \tau \approx \hat{\tau}_{orb}$, the time a particle takes to orbit a magnetic cell is comparable to the cell turnover time. In this case, according to our numerical calculations, particles appear to have the same diffusive behavior whether pitch angle scattering is turned on or off (red lines in Figure~\ref{fig:rundiffcellLgpa}). If $\hat \tau < \hat{\tau}_{orb}$, particles orbital motion is slower than cell turnover time (i.e. many cell turnovers for a single orbital motion) so that particles only feel the average cell structure with transverse size of $\approx \frac{\hat{L}}{2}\left( 1 + \frac{\hat{\lambda}}{\pi}\right)$. In this case particles diffuse only in the presence of pitch angle scattering (blue lines in Figure~\ref{fig:rundiffcellLgpa}). If $\hat \tau > \hat \tau_{orb}$, particle orbital motion is faster than the cell turnover time (i.e. many particle orbits for a single cell turnover). In this case, in the absence of pitch angle scattering, particles remain in the ballistic regime at the cell frequency, due to following the moving fieldlines. In contrast, pitch angle scattering makes the particles reach a steady diffusion regime, due to the additional cross fieldline motion (green lines in Figure~\ref{fig:rundiffcellLgpa}). 

The condition $\hat \tau \approx \hat \tau_{orb}$ imposes constrains on the magnetic cell geometry and the flow which underlie the time dependence. From eqns. (\ref{ExB}) and (\ref{eq:torb}), we see that $V/v$ is of order $\epsilon\hat\lambda/4$. For $v\sim c$ and $\hat\lambda = 0.25$, , $V$ ranges from 1.9 to 1900 km/s for the range of $\epsilon$ studied here. %
Thus, when magnetic chaos is present, charged particles appear to effectively move across fieldlines if $\hat \tau \approx \hat \tau_{orb}$, independently of pitch angle scattering.
%
 Additional evidence for the role of magnetic chaos in promoting diffusion comes from computations of $\hat \kappa_{\perp}$ at different $\hat\lambda$. We found that the diffusion coefficient of particles 
 is highest for the value of $\hat \lambda$ at which the average Lyapunov exponent is maximized. This is shown in Figure~\ref{fig:rundiffcellLgpb} in the case with $\hat \tau_{sca}\approx \hat \tau_{orb}$. In the discussion that follows, we consider the GP field structure case under the condition $\hat \tau \approx \hat \tau_{orb}$ and we choose the maximal value of $\hat \lambda \approx 0.25$ in our numerical calculation.

\begin{figure}[!ht]
\begin{center}
\includegraphics[width=\columnwidth]{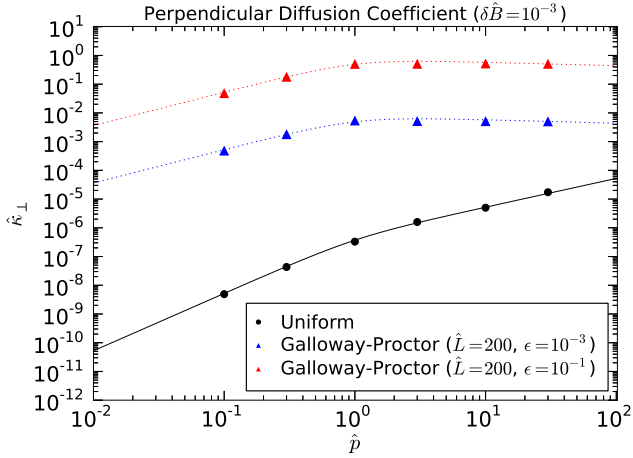}
\caption{Perpendicular diffusion coefficient in the GP magnetic geometry for magnetic perturbation $\delta \hat B$ = 10$^{-3}$, compared to the corresponding coefficients in the uniform magnetic field. The black continuous line is the quasilinear theory prediction eqn. (\ref{kperp}), and the dotted lines correspond to the fit result eqn. (\ref{eq:gpfit}). Numerical values can be found in Table \ref{tab:gpruns}.}
\label{fig:diffgp}
\end{center}
\end{figure}

Figures \ref{fig:rundiffgp2a} and \ref{fig:rundiffgp2b} shows the perpendicular running diffusion coefficient in the GP magnetic geometry with two different values of $\epsilon$ = 10$^{-3}$-10$^{-1}$, and pitch angle scattering simulated with $\delta \hat B$=10$^{-3}$. 
The figure shows that, similarly to the case of cellular structure (as shown in Figure~\ref{fig:rundiffcell}), transverse motion with respect to the guide field is ballistic until particles turn back in their orbital motion around the cell. Because fieldlines around a full cell perimeter are longer for smaller $\epsilon$, the ballistic behavior is clearly visible in Figure \ref{fig:rundiffgp2b}. On the other hand in the case of Figure \ref{fig:rundiffgp2a} it occurs at smaller times, not resolved in the plot. Similarly to the uniform structureless field geometry and the static cellular geometry, particle decorrelation is reached at scattering time $\hat \tau_{sca}$ although diffusion is turned on at or before $\hat\tau \approx \hat\tau_{orb}$, driven by the time dependence of the fieldlines. Similarly to the cellular magnetic field case, Figure \ref{fig:rundiffgp2c} shows the $\hat x$ coordinate distribution of 1000 particles in the GP magnetic field, on the $\hat z$=10$^7$ plane (integrated over all times). The time dependence of cellular magnetic field in the GP case facilitates a significantly larger cross-fieldline transport than the static cellular geometry. A dependency on the amplitude of the magnetic field structure is also observed. The resulting diffusion coefficients are enhanced compared to the case of cellular field structure, as shown in Figure~\ref{fig:diffgp}, compared to Figure \ref{fig:diffcell}. This means that the time dependence acts to amplify particle cross fieldline propagation, in accordance to Figure \ref{fig:rundiffgp2c}.

The electric field leads to stochastic
energization of the particles, but we find it to be small and do not discuss it further here.

%
%
%
%

We fitted the transverse diffusion coefficient in the GP magnetic structure case as in eqn. (\ref{eq:cellfit}), using the linearized function ln
\begin{equation}
\hat \kappa^{GP}_{\perp} = C\, \epsilon^{a_{\epsilon}}\, \hat p^{a_p}\, (1+\hat p^2)^{a_{p2}},
\label{eq:gpfit}
\end{equation}
where, in this case $\delta \hat B$ = 10$^{-3}$ and $\hat L$ = 200. The fit results are summarized in Table~\ref{tab:gpfit}, and shown with a dotted line in Figure \ref{fig:diffgp}. The errors are estimated using the covariance matrix.
%
\begin{table}[!ht]
\caption{\label{tab:gpfit} Fit parameters for\\GP magnetic structure}
\centering
\begin{tabular}{ c r }
\hline
parameter        & value($\pm$ error) \\
\hline
$\ln(C)$            & 2.06 $\pm$ 0.13 \\
$a_{\epsilon}$  & 1.01 $\pm$ 0.02 \\
$a_p$              & 1.15 $\pm$ 0.07 \\
$a_{p2}$         & -0.64 $\pm$ 0.05 \\
\hline
\end{tabular}
\end{table}
%

According to Table \ref{tab:gpfit} the diffusion coefficient is approximately $\hat \kappa_{\perp}^{GP} = 3.79\times 10^{-2}\hat L\epsilon\hat v\approx \hat L\epsilon\, \hat v$.
\begin{figure}[!ht]
\begin{center}
\subfloat[\label{fig:diffalla}]{%
  \includegraphics[width=\columnwidth]{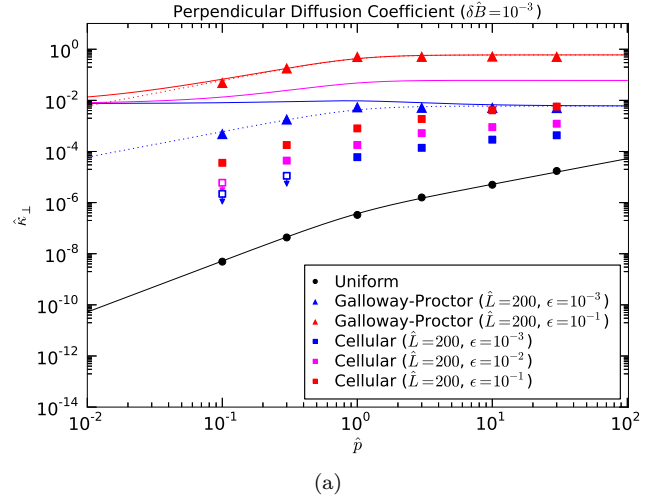}
  }

\subfloat[\label{fig:diffallb}]{%
  \includegraphics[width=\columnwidth]{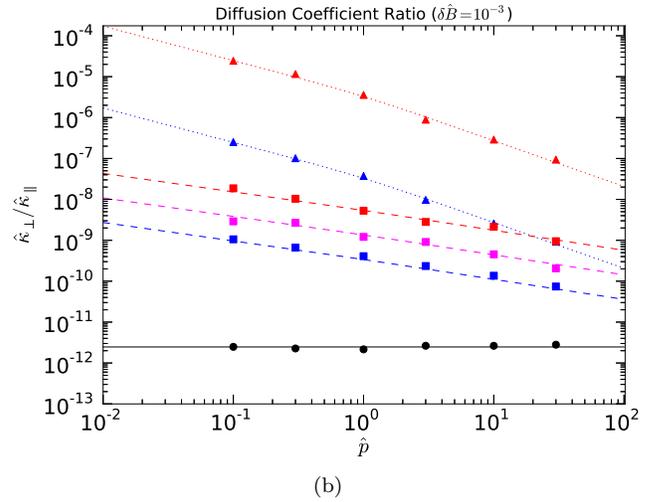}
  }
\caption{Summary of perpendicular diffusion coefficients $\hat \kappa_{\perp}$ (a) and of the ratio $\hat \kappa_{\perp}/\hat \kappa_{\parallel}$ (b) for magnetic perturbation geometries considered in this work with $\delta \hat B$ = 10$^{-3}$ (square and triangle symbols), compared to the corresponding coefficients in the uniform magnetic field (circle symbol). The black continuous line is the quasilinear theory prediction eqn. (\ref{kperp}), and it can be considered as the lower bound for perpendicular coefficient. The colored continuous lines in (a) represent the upper bounds eqn. (\ref{eq:Dmax}). The dotted lines in (a) correspond to the fit result eqn. (\ref{eq:gpfit}). The dotted and dashed lines in (b) corresponds to the fit results eqn. (\ref{eq:cellfit}) and (\ref{eq:gpfit}) respectively.}
\label{fig:diffall}
\end{center}
\end{figure}
%

\section{Discussion}\label{sec:discussion}

Figure~\ref{fig:diffalla} summarizes the perpendicular diffusion coefficients obtained with the uniform, cellular and GP magnetic fields with pitch angle scattering for $\delta \hat B$ = 10$^{-3}$. The enhancing effect of magnetic geometry and of time variability is evident from the results. Such effects are significant only in the perpendicular motion, as the parallel diffusion coefficient in the presence of weak ($\vert\nabla A\vert/B_0\sim\epsilon\ll1$) magnetic structure is generally similar to that in the uniform magnetic field (i.e. from quasilinear theory) to first order in $\epsilon$.
%
%
This can be seen also in of Figure~\ref{fig:diffallb}, where the diffusion coefficient ratio $\hat \kappa_{\perp}/\hat \kappa_{\parallel}$ is shown. These figures show that perpendicular diffusion is strongly amplified in the presence of a large scale geometric structure, and even more if time variations on the particle orbital time scale and/or chaos are present. The enhancement can be several orders of magnitudes compared to the uniform structureless magnetic field case, as long as particle gyroradius is smaller than the magnetic coherence length. On the other hand, perpendicular diffusion is still relatively small compared to parallel diffusion. An important and nontrivial finding is that the cross fieldline transport can indeed be described as diffusion, not superdiffusion or subdiffusion: $\langle (\Delta x_{\perp}^2)\rangle\propto t$ in the limit of large times. 

The figures show that the perpendicular diffusion coefficients in presence of a weak perpendicular magnetic perturbation converges to the uniform case with increasing particle gyroradius. Numerical calculation confirms such convergence. This is expected because when the gyroradius is larger that the cell size (or the largest perturbation scale in general), particles are influenced by the mean magnetic field within their gyration orbit, which is very close to the guide field, thus similar to a uniform field. On the other hand, at small gyroradii, particle motion is dominated by the larger spatial scale of magnetic perturbations. 

According to \cite{jokipii_1993, giacalone_1994, jones_1998}, if magnetic perturbations are limited to perpendicular structures with respect to the guide field, so the canonical momentum $P_{\parallel}$ along the guide field is conserved, particles will always be trapped in cells or magnetic vortices, thus limiting their perpendicular motion. However, in \cite{hauff_2010} it was found that 
perpendicular transport in purely 2D turbulent field can be diffusive if the particle gyroradius is $\gtrsim$ perpendicular coherence length of the field. In this case particles decorrelate on a time scale $\lesssim$ gyroperiod and are not bound to the magnetic fieldlines, but undergo stochastic motion. Although the magnetic fields considered here are invariant along $z$ (i.e. the direction of the guide field)
conservation of $P_{\parallel}$ is violated due to pitch angle scattering. Thus, the theorem of Jokipii and coauthors does not apply to our case.

We now estimate lower and upper bounds on $\hat \kappa_{\perp}$. Perpendicular transport across the uniform field, with diffusivity given by eqn. (\ref{kperp}), sets the lower bound, $\hat \kappa_{\perp,min}$. We estimate the upper
bound by replacing the step size $r_g$ in eqn. (\ref{kperp}) by the cell width $L/2$, and introducing an effective timescale $\hat\tau_{eff}$
%
\begin{equation}
\hat \kappa_{\perp, max} = \lim_{t\to \infty} \frac{\langle \Delta x(t)^2 \rangle}{2\Delta t} = \frac{L^2}{8\hat \tau_{eff}},
\label{eq:Dmax}
\end{equation}
where
the time scale $\hat \tau_{eff}$ is 
the smaller of the
 scattering time 
(i.e. the earliest time particles are released from a magnetic cell) and the particle eddy turnover time (i.e. either the particle orbital time in a static cellular field or the cell turnover time itself).
\begin{equation}
\hat \tau_{eff} = \left(\frac{1}{\hat \tau_{sca}} + \frac{1}{\hat t_{eddy}}\right)^{-1}.
\end{equation}
The upper bound on $\hat\kappa_{\perp}$, therefore, is
\begin{equation}
\hat \kappa_{\perp, max} = \frac{\hat L^2}{8\hat \tau_{sca}} + \hat \kappa_{\perp, eddy}.
\label{eq:Dmax2}
\end{equation}
Because for the GP field, as shown in Figure~\ref{fig:rundiffcellLgpa}, the highest diffusion coefficient is reached when the eddy turnover timescale $\hat\tau$ is approximately $ \hat \tau_{orb}$, independent of pitch angle scattering, it is possible to derive a general expression for the eddy diffusion coefficient
\begin{equation}\label{eddy}
\hat \kappa_{\perp, eddy} = \frac{\hat L^2}{8\hat \tau_{orb}} = \frac{\hat L\, \epsilon}{32}\, \frac{\hat p}{\sqrt{1+\hat p^2}},
\label{eq:kperpeddy}
\end{equation}
where we have used eqn. (\ref{eq:torb}). Our expression for $\kappa_{\perp,eddy}$ is proportional to $\hat L \epsilon v$ and correctly reproduces the fitted diffusion coefficient in the GP magnetic structure eqn. (\ref{eq:gpfit}), with the fit parameters of Table~\ref{tab:gpfit}. This is shown in Figure \ref{fig:diffalla} with dotted lines, while the continuous line is the maximum diffusion coefficient $\hat \kappa_{\perp, max}$. For a given cell size and magnetic perturbation strength eqn. (\ref{eq:Dmax2}) represents the upper bound in perpendicular coefficient. As shown earlier, changing the turnover time $\hat \tau$ and complexity parameter $\hat \lambda$ would decrease the diffusion coefficient. Note that $\kappa_{\perp,max}/\kappa_{\perp,min} > 1$ for $\hat p/\hat L < \epsilon/(16\pi)(\hat\delta B)^{-2}$. For all the runs considered here, $\kappa_{\perp,max}/\kappa_{\perp,min} > 1$ for $r_g/L = \hat p/\hat L$, as we require. However, if $\hat p/\hat L$ is sufficiently small, 
particles follow magnetic field lines and cross fieldline diffusion should revert to the uniform field case for sufficiently large scale structure.

Anything between the lower bound $\propto v \hat p$ and the upper bound $\propto \hat L \epsilon v$ comprise all the intermediate cases in which particles are sequentially trapped in a magnetic cell and released from it to migrate to another one. The exchange between the two conditions happens on the time scale of the particle decorrelation time. Our numerical calculations show in such cases the scaling of diffusion coefficient involves fractional powers of the main parameters. 

The cellular geometry case considered here is reminiscent of percolation, in which particles spread from one cell to another by the joint action of advection and diffusion. \cite{isichenko_1989} and \cite{isichenko_1992} considered transport of a passive scalar by a steady cellular flow with a single spatial scale in the presence of a small diffusivity $\kappa_0$. Denoting the eddy diffusivity of the flow by
$\kappa_t$ they found an effective diffusivity
\begin{equation}\label{kappa_eff}
\kappa_{eff}\propto\kappa_0^{3/13}\kappa_t^{10/13}.
\end{equation}
If we identify $\kappa_0$ with $r_g^2/\tau_{sca}\propto\hat p\hat v\delta\hat B^2$, the perpendicular diffusivity in a uniform magnetic field, and $\kappa_t$ with $\hat L^2/\hat\tau_{orb}$, we find that eqn. (\ref{kappa_eff}) predicts
\begin{equation}\label{kappa_isi}
\hat \kappa_{\perp}\propto\hat p^{0.23}\hat v\delta\hat B^{0.46}\epsilon^{0.77}\hat L^{0.77}
\end{equation}
which is quite different from the exponents given in Table 1. Bearing in mind that our determination of $\alpha_L$ is an upper limit, we see that we could improve the fit by including a weak dependence
on $\hat p/\hat L$, accounting for the fact that if the particle gyroradius includes a cell boundary, the particle can escape the cell more easily. For time dependent flows, $\kappa_{eff}$ can be 
close to the eddy value $\kappa_t$ (Isichenko 1992 and references therein), as we found for the GP magnetic field.

\subsection{Role of the Kubo Number}\label{ssec:kubo}

The Kubo number $\cal K$ is generally defined as the ratio of the distance a particle travels during an autocorrelation time over the correlation distance. 
For our problem,
\begin{equation}\label{K}
{\cal{K}}\sim\frac{\hat \tau_{sca}}{\hat \tau_{orb}}.
\end{equation}
For $\cal K$ smaller than 1 the distance particles travel before decorrelation is small, therefore there is no time to be trapped in a magnetic cell
and particles quickly become diffusive. If ${\cal K}\ge 1$, 
 particles are trapped in magnetic 
cells until decorrelation releases them so that they can pass to another 
cell. 
For the cellular magnetic field, $\cal K$ can be written as
\begin{equation}\label{Kcell}
{\cal K}\sim\frac{\mu}{2\pi}\frac{\epsilon\hat p}{\hat L}{\delta \hat B}^{-2}\sim\frac{800\mu\hat p}{\epsilon}\left(\frac{200}{\hat L}\right)\left(\frac{\delta \hat B}{10^{-3}}\right)^{-2},
\end{equation}
where the fiducial values $\hat L = 200$ and $\delta\hat B$ have been inserted in the last equality. For most of the parameter space we examined, ${\cal K}\gg 1$.

In previous studies of particle transport \citep{gruzinov_1990, isichenko_1992, neuer_2006a, hauff_2010} it was found that diffusion perpendicular to the mean magnetic field
scales with $\cal K$.
In \cite{hauff_2010}, a general expression for diffusion perpendicular to the mean field is proposed
\begin{equation}\label{Kscaling}
\hat \kappa_{\perp}\sim {\cal K}^{\gamma}\frac{\hat L^2}{\hat \tau_{sca}}\propto\epsilon^{\gamma}{\hat L}^{2-\gamma}\frac{\hat p^{\gamma}}{\sqrt{1+\hat p^2}}{\delta \hat B}^{2-2\gamma},
\end{equation}
where for ${\cal K} > 1$, $\gamma\sim 0.7$. The fractional power of this scaling law relates to the percolative behavior of particle transport of subsequent trapping and releasing from turbulent magnetic vortices. In this study we found that perpendicular transport in a simple periodic cellular magnetic field does not show the same behavior as advection/diffusion of a passive scalar in a flow with the same
simple cellular structure. Likewise,
we find that the Kubo number as defined in eqn. (\ref{K})
does not provide the correct scaling, as it does for turbulent magnetic field cases. For $\hat p = (0.1, 1, 10)$, we find that the scaling with $\epsilon$ suggests $\gamma\sim (0.61, 0.56, 0. 85)$
over 2 orders of magnitude in $\epsilon$. For $\hat p = 10$, the scaling with $\hat L$ suggests $\gamma\sim 1.0$ over 1 order of magnitude in $\hat L$. The scaling with $\hat p$ suggests an even larger
value of $\gamma$; $\gamma\sim 1.6$, for $0.1 < \hat p < 10$. We suspect that this larger value of $\gamma$ is due to finite gyroradius effects at the larger values of $\hat p$.
Perhaps a more general definition of $\cal K$ can provide the correct scaling in a larger set of situations.

\subsection{Application to the Fermi Bubbles}

Observations from Fermi-LAT reveal two large bubbles that extend 50$^{\circ}$ above and below the galactic center, with a width of about 40$^{\circ}$ in longitude \citep{su_2010}.
Simulations of the Fermi Bubbles as overpressured structures driven by bursts of AGN activity into the ambient interstellar medium were developed by \citet{guo_2012a}, \cite{guo_2012b}, \cite{yang_2012}, and \cite{yang_2013}.
It was found that if  diffusion perpendicular to the magnetic field is completely suppressed, the resulting cosmic ray surface density profile is sharper,
and in better agreement with observations, than
if diffusion is isotropic. Anisotropic diffusion has a strong effect because the  magnetic field tends to be draped over the expanding bubble, giving it a preferred orientation. Here, we  estimate whether complete suppression of perpendicular diffusion is consistent with the results of the present paper,
given the
magnetic and flow geometry in the simulations\footnote{Note that we are not claiming that the episodic AGN model described in \cite{yang_2012} is the only
viable model; we are only testings its treatment of cosmic ray diffusion for self consistency.}.
  
\begin{figure}[!ht]
\begin{center}
\includegraphics[width=\columnwidth]{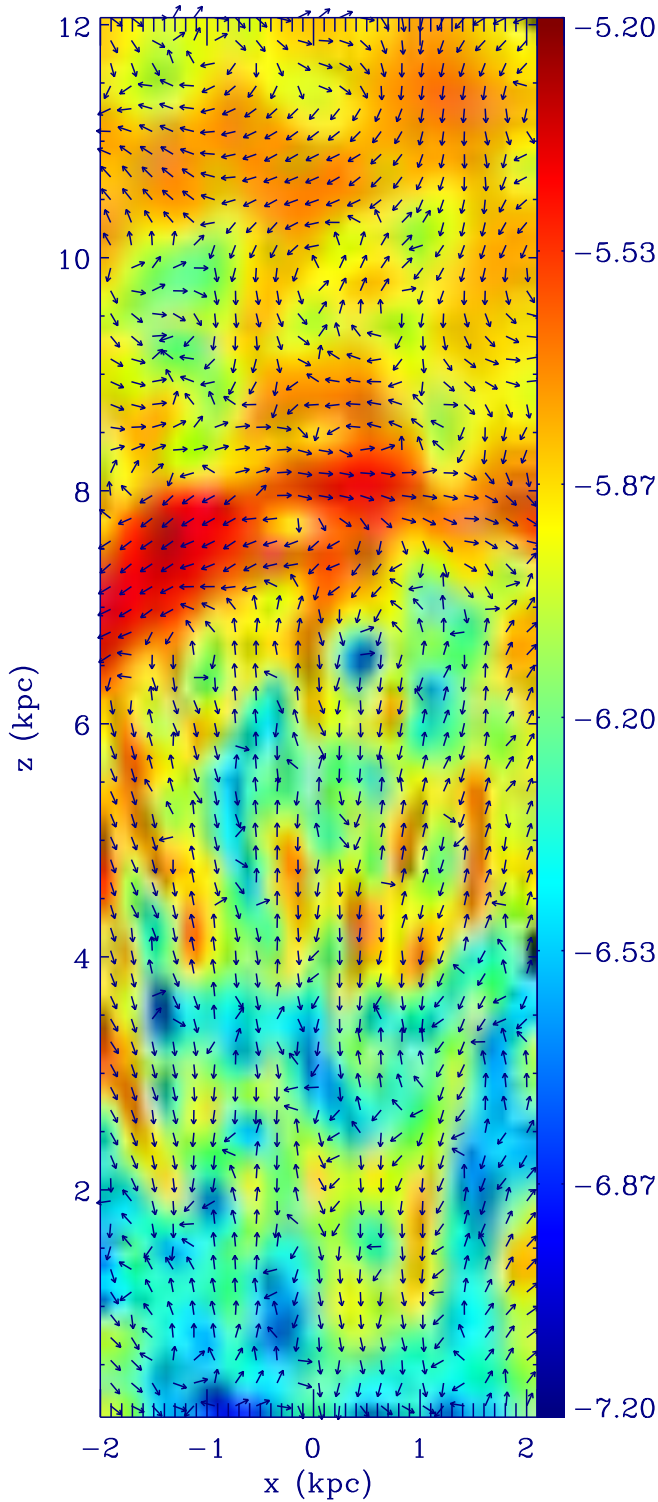}
\caption{Detailed slice in simulation coordinates of magnetic field morphology in a model Fermi Bubble \citep{yang_2012} where field strength (in logarithmic scale in units of $\mu$G) is represented in the color scale and  projected direction by arrows. Figure courtesy of H.-Y. K. Yang.}
\label{fig:fbubble}
\end{center}
\end{figure}

Figure \ref{fig:fbubble} shows a 4 $\times$ 12 square kpc portion of one of the simulations described in \cite{yang_2012}.
 The magnetic field morphology and strength are indicated by arrows and by colors, respectively. 
There is clearly a transition layer between the bubble interior and the ambient medium, located roughly at $z\sim 7.5 - 8.5$ kpc, in which the magnetic field is roughly tangent to the expanding front. It is assumed in the simulation that cosmic rays diffuse parallel to the magnetic field with diffusivity $\kappa_{\parallel} = 4\times 10^{28}$ cm$^2$ s$^{-1}$, corresponding to a mean free path of about 1.3 pc and a scattering time $\tau_{sca}\sim 10^8$. 
%

 Cosmic rays are prevented from leaking across the transition layer between the bubble and the ambient medium if the perpendicular diffusion coefficient $\kappa_{\perp}$, front speed $V_f$, and layer thickness $D$ satisfy the inequality $\kappa_{\perp} < DV_f$. Since the magnetic layer is evidently tangent to the front, it is reasonable to assume that the perpendicular eddy size $L_{\perp}$ is small compared to the magnetic correlation length along the front, and that the component tangent to the front dominates. We can then take the parametric dependences in eqn. (\ref{eddy}) as an upper bound on the diffusivity, such that leaking is prohibited if
 \begin{equation}\label{noleak}
 \epsilon L_{\perp}c < DV_f,
 \end{equation}
 where we assume $v\sim c$.
 
Unfortunately, the global simulations at hand do not resolve magnetic field structure below 100 pc in size, so we cannot measure $\epsilon$ and $L_{\perp}$ directly from the models. Likewise, the computed transition layer thickness may be an overestimate. However, eqn. (\ref{noleak}) seems easy to satisfy. Taking $V_f/c \sim 0.02$ (Yang et al. 2012) and $D\sim 100$ pc (less than the simulated
 layer thickness) we see that the right hand side of the inequality (\ref{noleak}) is $1.8\times 10^{29}$ cm$^2$ s$^{-1}$. This is several times larger than $\kappa_{\parallel}$, while we have found 
 $\kappa_{\perp}/\kappa_{\parallel}\ll 1$. In order to apply our theory, $L_{\perp}/\epsilon$ must be less than the scattering mean free path $\lambda_{\parallel}\sim 1 $pc. If we set $\epsilon = 0.1$ and $L_{\perp} = 0.1$pc, the left hand side of the inequality (\ref{noleak}) is $10^{27}$ cm$^2$ s$^{-1}$. Thus, it is plausible that the cosmic rays are well confined to the bubbles.

 \section{Conclusions}\label{sec:conclusions}
 
 In this paper we have investigated the transport of cosmic rays across magnetic fields. In particular, we considered the interplay between magnetic geometry and cross fieldline transport by pitch angle scattering. For simplicity, we chose simple, single scale 2D cellular magnetic geometries superimposed on a guide magnetic field. One case is static, with integrable fieldlines, and the other is periodic in time, with chaotic fieldlines. The scattering is assumed to be due to gyroscale magnetic turbulence, while the cell size is well above the gyroradius. Our study complements the large body of work on transport across {\textit{mean}} magnetic fields. 
 
 We showed that the running spatial diffusion coefficients $D(t)$ defined in eqn. (\ref{runningD}) approach steady values $\kappa$ over time, indicating diffusive (as opposed to subdiffusive or superdiffusive) behavior. We argued that due to the bounded horizontal excursions of the fieldlines chosen for our study, $\kappa_{\perp}$ represents true cross fieldline transport. Cellular structure enhances transport by several orders of magnitude as compared to transport in a uniform field. We found that fields which vary on a timescale similar to the timescale on which particles orbit the cells have fast transport even without pitch angle scattering. But in all the cases we considered, perpendicular diffusion is much slower than parallel diffusion. We derived upper and lower bounds on the diffusivity and compared our results to those obtained for advection-diffusion of passive scalars, and in fully turbulent magnetic fields, but did not find the same scalings. We did, however, apply our results to particle confinement in magnetically bounded structures such as the Fermi Bubbles, and argued that escape by cross fieldline transport would be slow.
 
 In reality, cosmic rays encounter magnetic turbulence on a vast range of scales. Each particle will be affected more by some scales than others, but the interplay of multiple scales is sure to give rise to interesting effects not encountered here. We will take up some of these effects in future work.
 
 \acknowledgments
We are happy to acknowledge support by NSF grants AST 0907837 and PHY 0821899 (to the Center for Magnetic Self-Organization), and useful discussions with John Finn, Frank Jenko and Reinhard Schlickeiser. Many thanks to H.-Y. Karen Yang and Mateusz Ruszkowski for providing 
results
 on magnetic field modeling of the Fermi Bubbles. PD wish to thank the Wisconsin IceCube Particle Astrophysics Center (WIPAC). The authors also thank Dr. Pierre Barge for useful comments and feedback.


\begin{appendix}

\section{Numerical Approach and Accuracy}
\label{asec:implement}

We solved the relativistic equation of motion~(\ref{dimlessmomentum}) using numerical integration methods and adaptive time step algorithms provided by the GNU Scientific Libraries \citep{gsl_2009}. This gives the flexibility to assess the solution accuracy by comparing different integration methods. For this study the explicit 4$^{th}$ order Runge-Kutta (RK4) integration method was used for all the generated datasets. The implicit Gaussian 2$^{nd}$ order Runge-Kutta (RK2), also known as implicit mid-point method, was used for a selected number of datasets to test the relative accuracy. 
Adaptive time stepping was implemented by
advancing the solution using a given time step and two halves of it (step doubling method) and by comparing the difference in the system states $\by(t)$ to the specified absolute and relative error levels $\epsilon_{abs}$ and $\epsilon_{rel}$. The default error levels were chosen to be $\epsilon_{err}\equiv \epsilon_{abs} = \epsilon_{rel} = 10^{-10}$. Variations of this value ($\epsilon_{err} = 10^{-12}, 10^{-14}$) were used to estimate the accuracy of the solution as described below. 


In the 2.5D cases studied here, where $z$ is an ignorable coordinate, 
it is easy to verify, from eqn. (\ref{dimlessmomentum}), that the component of canonical momentum parallel to the guide magnetic field $p_z + \frac{q}{c}A_z\rightarrow \hat p_z + \hat A_z \equiv P_z$, is conserved, making it possible to decrease the dimensionality of the system from six to four, and solve eqn. (\ref{dimlessmomentum}) on the (x,y) plane only. Such reduction is not possible
if pitch angle scattering is imposed.
Reducing the dimensionality of the particle motion equation improves the accuracy of the solution in most circumstances. 
Comparison of the 4D and 6D solutions thus tests the accuracy of the full 6D treatment required when pitch angle scattering is present.




We tested our numerical results in three different ways. First we checked how well energy is conserved for orbits computed by both explicit and implicit integration stepping methods in both 4D and 6D, in the latter case with and without pitch angle scattering. Then we checked how well canonical momentum is conserved for orbits computed in 6D without pitch angle scattering. And finally we also compared the morphology of a few orbits computed in 4D and 6D with both the implicit and explicit stepping methods to orbits computed with the NDSolve feature of Mathematica. 

The uniform magnetic field is a benchmark case for testing the integration accuracy. A constant relative rounding error in particle momentum of about
$\frac{d\hat p}{\hat p}\approx 10^{-7}$-$10^{-6}$
is 
manifested as a spread around the average momentum. In addition to this spread a slow energy dissipation is also observed,
With the explicit RK4 integration method, the total energy loss rate is estimated to be about
%
$d\hat p/ds \approx -10^{-14}$-$10^{-12}$,
depending on the particle pitch angle. For parallel propagation
the energy dissipation is at the lower end of the range, while for perpendicular propagation
it is at the upper end, which is the rate of perpendicular energy loss for an arbitrary pitch angle. These 
estimates are found to be fairly independent of particle energy, and also to be consistent with the rate observed with more structured static magnetic fields, such as the cellular configuration. This means that with the explicit RK4 integration method, an energy loss of 1\% is reached at integration time 
$\approx 10^9$-$10^{11}$ for $\hat p = 0.1$ and $
\approx 10^{11}$-$10^{13}$ for $\hat p = 10$, depending on pitch angle.

Energy conservation improves if the adaptive step size error level $\epsilon_{err}$ is reduced. If $\epsilon_{err} = 10^{-12}$ (i.e. two orders of magnitude smaller than the default value for this study), no significant energy dissipation is observed up to integration time $s = 10^{10}$, while the rounding error remains at the same level. 
This accuracy level is reached with a penalty of about three times longer computation times. Using the implicit RK2 integration method (with $\epsilon_{err} = 10^{-10}$), accuracy is greatly improved, but computation time is increased by at least a factor of 6. We generally integrate to $s=10^9$, so explicit RK4 with $\epsilon_{err} = 10^{-10}$ is sufficient to keep integration accuracy at an acceptable level for the present study. 

Accuracy of energy conservation in the presence of pitch angle scattering was also found to be $d\hat p/ds \approx 10^{-12}$-$10^{-11}$, independently of particle pitch angle, and magnetic field geometry. Since scattering processes are simulated so as to conserve particle energy (see \S\ref{ssec:sca}), the slight worsening of energy conservation compared to the case without scattering can be attributed to repeated application of the rotation matrices we use to simulate pitch angle scattering.

Conservation of the $z$-component of canonical momentum in the 6D formulation
was tested with the cellular magnetic geometry perturbation case, that conserves $P_z \equiv p_z+\frac{q}{c}A_z$. Typically, for 
nearly parallel propagation the canonical momentum is conserved although it has a relative spread $\frac{dP_z}{P_z}\approx 10^{-6}$-$10^{-5}$, relatively comparable with rounding error. For 
nearly perpendicular propagation, however, dissipation is also observed with a rate $\frac{dP_z}{ds}\approx -10^{-12}$-$10^{-11}$. 


In this study a coupling between transverse and parallel diffusion was not observed within the parameter space investigated here. The divergence of magnetic field lines in large scale turbulent interstellar magnetic field may induce a coupling between them \citep{barge_1984}.

Finally, a qualitative comparison of particle trajectories calculated in a cellular magnetic field geometry, with explicit RK4 and implicit RK2 integration methods in 6D and 4D, and with the solutions obtained with Mathematica, was carried out. 
The trajectory geometries were identical to each other within the assessed integration accuracy in the case of regular trajectories. When a trajectory appears to 
be chaotic in nature (which occurs for specific initial conditions), chaos \footnote{when chaotic trajectories are found with the different integrators and dimensionality, the actual trajectories are different from each other, while the general characteristics are reproduced, as expected for chaos.} is observed for all the tested integration methods and for the 4D and 6D formulations. We find chaos only for high momentum particles with $\mu\sim 1$ and gyroradii comparable to the magnetic cell size. Our model, which is intended to follow propagation in large scale resolved magnetic structure with pitch angle scattering by small scale unresolved structure, is not well motivated for these high momentum particles, so we do not study them further here.

\section{Datasets}

\begin{ruledtabular} 
\begin{table*}[!ht]
\caption{\label{tab:uniruns} List of particle integration datasets for the Uniform Magnetic Field.}
\begin{tabular*}{\textwidth}{c c c c c}

$\hat p$ & $\delta \hat B$ & $\hat \tau_{sca}$       &  $\hat\kappa_{\parallel}$ & $\hat\kappa_{\perp}$ \\
\hline
$0.1$       & $10^{-3}$ &  $6.4\times10^5$ & $2.0\times10^3$ & $4.9\times10^{-9}$  \\
$0.3$       & $10^{-3}$ &  $6.6\times10^5$ & $1.9\times10^4$ & $4.3\times10^{-8}$  \\
$1$          & $10^{-3}$ &  $9.0\times10^5$ & $1.5\times10^5$ & $3.3\times10^{-7}$  \\
$3$          & $10^{-3}$ &  $2.0\times10^6$ & $6.0\times10^5$ & $1.6\times10^{-6}$  \\
$10$        & $10^{-3}$ &  $6.4\times10^6$ & $1.9\times10^6$ & $5.0\times10^{-6}$  \\
$30$        & $10^{-3}$ &  $1.9\times10^7$ & $6.2\times10^6$ & $1.7\times10^{-5}$  \\\hline
$1$          & $10^{-2}$ &  $9.0\times10^3$ & $1.6\times10^3$ & $3.7\times10^{-5}$  \\
$30$        & $10^{-2}$ &  $1.9\times10^5$ & $6.7\times10^4$ & $1.6\times10^{-3}$  \\

\end{tabular*}
\end{table*}
\end{ruledtabular}   

\begin{ruledtabular}  
\begin{table*}[!ht]
\caption{\label{tab:cellruns} List of particle integration datasets for the Cellular Magnetic Field. Note that when diffusion regime (i.e. convergence of the running diffusion coefficient) was not reached within integration time of 10$^9$ the coefficient is indicated with the symbol $<$.}
\begin{tabular*}{\textwidth}{c c c c c c c c}

$\hat p$ & $\delta \hat B$ & $\hat L$ & $\epsilon$ & $\hat \tau_{sca}$       & $\hat \tau_{orb}$     & $\hat\kappa_{\parallel}$ & $\hat\kappa_{\perp}$ \\
\hline
$0.1$       & $10^{-3}$ & 200        & $10^{-3}$  & $6.4\times10^5$ & $8.0\times10^6$ & $2.0\times10^3$ & $<2.2\times10^{-6}$  \\
                & $10^{-3}$ & 200        & $10^{-2}$  & $6.4\times10^5$ & $8.0\times10^5$ & $2.0\times10^3$ & $<6.0\times10^{-6}$  \\
                & $10^{-3}$ & 200        & $10^{-1}$  & $6.4\times10^5$ & $8.0\times10^4$ & $2.0\times10^3$ & $3.6\times10^{-5}$  \\ \hline
$0.3$       & $10^{-3}$ & 200        & $10^{-3}$  & $6.6\times10^5$ & $2.8\times10^6$ & $1.7\times10^4$ & $<1.1\times10^{-5}$  \\
                & $10^{-3}$ & 200        & $10^{-2}$  & $6.6\times10^5$ & $2.8\times10^5$ & $1.6\times10^4$ & $4.4\times10^{-5}$  \\
                & $10^{-3}$ & 200        & $10^{-1}$  & $6.6\times10^5$ & $2.8\times10^4$ & $1.7\times10^4$ & $1.8\times10^{-4}$  \\\hline
$1$          & $10^{-3}$ & 200        & $10^{-3}$  & $9.0\times10^5$ & $1.1\times10^6$ & $1.5\times10^5$ & $6.0\times10^{-5}$  \\
                & $10^{-3}$ & 200        & $10^{-2}$  & $9.0\times10^5$ & $1.1\times10^5$ & $1.5\times10^5$ & $1.8\times10^{-4}$  \\
                & $10^{-3}$ & 200        & $10^{-1}$  & $9.0\times10^5$ & $1.1\times10^4$ & $1.5\times10^5$ & $8.0\times10^{-4}$  \\\hline
$3$          & $10^{-3}$ & 200        & $10^{-3}$  & $2.0\times10^6$ & $8.4\times10^5$ & $5.9\times10^5$ & $1.4\times10^{-4}$  \\
                & $10^{-3}$ & 200        & $10^{-2}$  & $2.0\times10^6$ & $8.4\times10^4$ & $5.7\times10^5$ & $5.2\times10^{-4}$  \\
                & $10^{-3}$ & 200        & $10^{-1}$  & $2.0\times10^6$ & $8.4\times10^3$ & $6.7\times10^5$ & $1.9\times10^{-3}$  \\\hline
$10$        & $10^{-3}$ & 200        & $10^{-3}$  & $6.4\times10^6$ & $8.0\times10^5$ & $2.1\times10^6$ & $2.9\times10^{-4}$  \\
                & $10^{-3}$ & 200        & $10^{-2}$  & $6.4\times10^6$ & $8.0\times10^4$ & $2.0\times10^6$ & $9.0\times10^{-4}$  \\
                & $10^{-3}$ & 200        & $10^{-1}$  & $6.4\times10^6$ & $8.0\times10^3$ & $2.0\times10^6$ & $4.1\times10^{-3}$  \\\hline
$30$        & $10^{-3}$ & 200        & $10^{-3}$  & $1.9\times10^7$ & $8.0\times10^5$ & $5.8\times10^6$ & $4.3\times10^{-4}$  \\
                & $10^{-3}$ & 200        & $10^{-2}$  & $1.9\times10^7$ & $8.0\times10^4$ & $5.9\times10^6$ & $1.2\times10^{-3}$  \\
                & $10^{-3}$ & 200        & $10^{-1}$  & $1.9\times10^7$ & $8.0\times10^3$ & $6.0\times10^6$ & $5.7\times10^{-3}$  \\\hline
\hline
$1$          & $10^{-2}$ & 200        & $10^{-3}$  & $9.0\times10^3$ & $1.1\times10^6$ & $1.5\times10^3$ & $1.5\times10^{-4}$  \\
                & $10^{-2}$ & 200        & $10^{-1}$  & $9.0\times10^3$ & $1.1\times10^4$ & $1.5\times10^3$ & $5.5\times10^{-3}$  \\\hline
$30$        & $10^{-2}$ & 200        & $10^{-3}$  & $1.9\times10^5$ & $8.0\times10^5$ & $6.3\times10^4$ & $3.9\times10^{-3}$  \\
                & $10^{-2}$ & 200        & $10^{-1}$  & $1.9\times10^5$ & $8.0\times10^3$ & $6.1\times10^4$ & $4.7\times10^{-2}$  \\\hline
\hline
$30$        & $10^{-1}$ & 200        & $10^{-3}$  & $1.9\times10^3$ & $8.0\times10^5$ & $6.1\times10^2$ & $4.9\times10^{-3}$  \\
                & $10^{-1}$ & 200        & $10^{-1}$  & $1.9\times10^3$ & $8.0\times10^3$ & $5.9\times10^2$ & $1.7\times10^{-1}$  \\
                
\end{tabular*}
\end{table*}
\end{ruledtabular}   

\begin{ruledtabular}  
\begin{table*}[!ht]
\caption{\label{tab:gpruns} List of particle integration datasets for the GP Magnetic Field.}
\begin{tabular*}{\textwidth}{c c c c c c c c c c}

$\hat p$ & $\delta \hat B$ & $\hat L$ & $\epsilon$ & $\hat \tau_{sca}$       & $\hat \tau_{orb}$     & $\hat \tau$               & $\hat \lambda$ & $\hat\kappa_{\parallel}$ & $\hat\kappa_{\perp}$ \\
\hline
$0.1$       & $10^{-3}$ & 200        & $10^{-3}$  & $6.4\times10^5$ & $8.0\times10^6$ & $10^7$             & $0.25$              & $2.0\times10^3$ & $5.1\times10^{-4}$  \\
                & $10^{-3}$ & 200        & $10^{-1}$  & $6.4\times10^5$ & $8.0\times10^4$ & $10^5$             & $0.25$              & $2.0\times10^3$ & $5.1\times10^{-2}$  \\ \hline
$0.3$       & $10^{-3}$ & 200        & $10^{-3}$  & $6.6\times10^5$ & $2.8\times10^6$ & $3\times10^6$ & $0.25$              & $1.9\times10^4$ & $1.9\times10^{-3}$  \\
                & $10^{-3}$ & 200        & $10^{-1}$  & $6.6\times10^5$ & $2.8\times10^4$ & $3\times10^4$ & $0.25$              & $1.6\times10^4$ & $1.9\times10^{-1}$  \\\hline
$1$          & $10^{-3}$ & 200        & $10^{-3}$  & $9.0\times10^5$ & $1.1\times10^6$ & $10^6$            & $0.25$              & $1.5\times10^5$ & $5.7\times10^{-3}$  \\
                & $10^{-3}$ & 200        & $10^{-1}$  & $9.0\times10^5$ & $1.1\times10^4$ & $10^4$            & $0.25$              & $1.5\times10^5$ & $5.3\times10^{-1}$  \\\hline
$3$          & $10^{-3}$ & 200        & $10^{-3}$  & $2.0\times10^6$ & $8.4\times10^5$ & $10^6$            & $0.25$              & $5.6\times10^5$ & $5.4\times10^{-3}$  \\
                & $10^{-3}$ & 200        & $10^{-1}$  & $2.0\times10^6$ & $8.4\times10^3$ & $10^4$            & $0.25$              & $6.0\times10^5$ & $5.4\times10^{-1}$  \\\hline
$10$        & $10^{-3}$ & 200        & $10^{-3}$  & $6.4\times10^6$ & $8.0\times10^5$ & $10^6$            & $0.25$              & $2.1\times10^6$ & $5.4\times10^{-3}$  \\
                & $10^{-3}$ & 200        & $10^{-3}$  & $6.4\times10^6$ & $8.0\times10^5$ & $10^8$            & $0.1$                & $2.1\times10^6$ & $4.4\times10^{-4}$  \\
                & $10^{-3}$ & 200        & $10^{-3}$  & $6.4\times10^6$ & $8.0\times10^5$ & $10^6$            & $0.1$                & $2.3\times10^6$ & $4.0\times10^{-3}$  \\
                & $10^{-3}$ & 200        & $10^{-3}$  & $6.4\times10^6$ & $8.0\times10^5$ & $10^4$            & $0.1$                & $1.8\times10^5$ & $9.5\times10^{-4}$  \\
                & $10^{-3}$ & 200        & $10^{-1}$  & $6.4\times10^6$ & $8.0\times10^3$ & $10^4$            & $0.25$              & $1.9\times10^6$ & $5.5\times10^{-1}$  \\\hline
$30$        & $10^{-3}$ & 200        & $10^{-3}$  & $1.9\times10^7$ & $8.0\times10^5$ & $10^6$            & $0.25$              & $5.8\times10^6$ & $5.3\times10^{-3}$  \\
                & $10^{-3}$ & 200        & $10^{-1}$  & $1.9\times10^7$ & $8.0\times10^3$ & $10^4$            & $0.25$              & $5.7\times10^6$ & $5.4\times10^{-1}$  \\\hline
\hline
$10$        &       $0$    & 200        & $10^{-3}$  & $\infty$                & $8.0\times10^5$ & $10^6$            & $0.1$                & --                         & $5.0\times10^{-3}$  \\
                &       $0$    & 200        & $10^{-3}$  & $\infty$                & $8.0\times10^5$ & $10^6$            & $0.25$              & --                         & $6.6\times10^{-3}$  \\
                &       $0$    & 200        & $10^{-3}$  & $\infty$                & $8.0\times10^5$ & $10^6$            & $0.5$                & --                         & $6.0\times10^{-3}$  \\
                &       $0$    & 200        & $10^{-3}$  & $\infty$                & $8.0\times10^5$ & $10^6$            & $1$                   & --                         & $3.4\times10^{-3}$  \\

\end{tabular*}
\end{table*}
\end{ruledtabular}   

\end{appendix}

\clearpage


\end{document}